\documentclass[preprint2]{aastex63}

\usepackage{natbib}
\usepackage{wasysym}
\usepackage{txfonts}
\def\EBV{\mbox{E$_{\rm B-V}$}}

\def\AV{\mbox{A$_{\rm V}$}}

\def\WCO{\mbox{${\rm W}_{\rm CO}$}}         
\def\ACO{\mbox{${\rm A}_{\rm CO}$}}         
\def\WCS{\mbox{${\rm W}_{\rm CS}$}}         
\def\WHCN{\mbox{${\rm W}_{\rm HCN}$}}

\def\Wth{\mbox{${\rm W}_{\rm 13}$}}         
\def\Ath{\mbox{${\rm A}_{\rm 13}$}}         
\def\Wei{\mbox{${\rm W}_{\rm 18}$}}

\def\nH2{\mbox{${\rm n}(\HH$)}}
\def\enH2{\mbox{$n_{(\HH$)}}}

\def\pccc{~{\rm cm}^{-3}} 
 
\def\pcc {~{\rm cm}^{-2}}

\def\Tstar {\mbox{${\rm T}_{\rm r}^*$}}
\def\Tsub#1 {\mbox{$T_#1$}}
\def\TK  {\Tsub K }
\def\TB  {\Tsub B }

\def\fH2{\mbox{f$_\HH$}}
\def\mfH2{\mbox{$<{\rm f}_\HH>$}}

 \def\arcmin{\mbox{$^{\prime}$}}

\def\degr{$^{\rm o}$}
\def\p{\mbox{$^+$}}
\def\z{\mbox{${^0}$}}

\def\cch{\mbox{C$_2$H}}

\def\h13cop{\mbox{{H$^{13}$CO\p}}}

\def\c3h2{\mbox{C$_3$H$_2$}}
 
 \def\R0{R$_0$}
\def\G0{G$_0$}

\def\ddeg{{}^\circ\kern-.1em}

\def\kms{\mbox{km\,s$^{-1}$}}

\def\bll{BL Lac}

\def\E#1 {$10^{#1}$}
\def\E#1 {E{#1}}
\def\P#1,{$\nH2\TK~=~#1\times~10^4\pccc$~K}
\def\ec#1,#2,#3,{#1\,(#2)\E{#3}}

\def\zoph{$\zeta$ Oph}

\def\H3{\mbox{H$_3$}}

\def\RH2{\mbox{R$_{\rm G}$}}
\def\GH2{\mbox{$\Gamma_{\HH}$}}
\def\g13{\mbox{g$_{13}$}}

\def\kHeH2{\mbox{$k_{ He-\HH}$}}

\def\tim#1,#2{\mbox{{$#1\times10^{#2}$}}}

\newcommand{\emm}[1]{\ensuremath{#1}}   
\newcommand{\emr}[1]{\emm{\mathrm{#1}}} 

\newcommand{\hcop}{\emr{HCO^+}} 
\def\Whcop{W$_{\hcop}$}
 
\newcommand{\HH}{\emr{{\rm H}_2}}
\newcommand{\cotw}{\emr{^{12}CO}}
\renewcommand{\coth}{\emr{^{13}CO}}

\newcommand{\coei}{\emr{C^{18}O}}

\newcommand{\N}[1]{\emr{N_{#1}}}

\newcommand{\X}[1]{\emm{X_\emr{#1}}}

\newcommand{\f}[1]{\emm{f_\emr{#1}}}

\newcommand{\Kkms}{\emr{\,K\,km\,s^{-1}}}

\shorttitle{Emission from diffuse molecular gas}
\shortauthors{H. S. Liszt }

\begin{document}

\date{generated \today}


\title{CO emission and CO hotspots in diffuse molecular gas}


\author{Harvey S. Liszt}
\affil{National Radio Astronomy Observatory \\
            520 Edgemont Road,
           Charlottesville, VA,
           22903-2475}
\email{hliszt@nrao.edu}



\begin{abstract}

We observed $\lambda$3mm \cotw, \coth, \coei, \hcop, HCN and CS 
emission from diffuse molecular gas along sightlines with \EBV\ $\approx$ 
0.1 - 1 mag. Directions were mostly chosen for their proximity to sightlines
toward background mm-wave continuum sources studied in \hcop\ absorption, 
at positions where maps of \cotw\ at 1\arcmin\ resolution showed surprisingly 
bright integrated CO J=1-0 emission \WCO\ = 5-12 K-\kms\ but we also observed 
in L121 near \zoph. Coherence emerges when data are considered over a broad range of 
\cotw\ and \coth\ brightness. \WCO/\Wth\ and N(\cotw)/N(\coth) are 20-40 for \WCO\ 
$\la 5$ K-\kms\ and N(CO) $\la 5\times 10^{15}\pcc$, increasing with much scatter 
for larger \WCO\ or N(CO). N(\coth)/N(\coei) $> 20-40$ ($3\sigma$) vs. an intrinsic 
ratio $^{13}$C/$^{18}$O = 8.4, from a combination of selective photodissociation 
and enhancement of \coth.  The observations are understandable if \cotw\ forms 
from the thermal recombination of \hcop\ with electrons, after which the observed 
\coth\ forms via endothermic carbon isotope exchange with $^{13}$C\p.  \WCS/\WCO\ 
increases abruptly for  \WCO\ $\ga 10$ K-\kms\ and \WCS/\Whcop\ is 
bimodal, showing two branches having N(CS)/N(\hcop) $\approx 5$ and 1.25.
Because CO formation and \hcop\ excitation both involve collisions between \hcop\ 
and ambient electrons, comparison of the CO and \hcop\ emission shows that 
the CO hotspots are small regions of enhanced N(CO) occupying only a small fraction 
of the column density of the medium in which they are embedded.  \hcop/CO and HCN/CO 
brightness ratios are 1-2\% with obvious implications for determinations of 
the true dense gas fraction.

\end{abstract}



\section{Introduction}

The sky viewed in CO J=1-0 $\lambda$2.6mm emission \citep{DamHar+01} is usually 
understood as the map of molecular gas, showing the locations of the Galactic 
molecular clouds.  CO emission and molecular gas are synonymous, 
even to the extent that a linear scaling of the integrated CO brightness \WCO\ 
(with units of K-\kms), with appropriate caveats, gives N(\HH) via the CO-\HH\ 
conversion factor \citep{BolWol+13}.

Interpretation of the CO sky map  is complicated by the presence of emission arising 
in 
partially molecular gas having a significant admixture of atomic hydrogen, 
small CO column densities and small CO abundances. CO lines from this so-called 
diffuse or diffuse molecular gas can be quite bright at \EBV\ $\la 0.3$ mag, 
\WCO\ $\simeq$ 
5-10 K-\kms, mimicking emission from traditional dark clouds 
\citep{LisPet+09,LisPet12}.   For example, the mean reddening toward the catalog 
positions of the MBM high-latitude molecular cloud complexes \citep{MagBli+85} 
is only 0.28$\pm$0.24 mag. Free gas-phase carbon is predominantly in the form
of C\p, not CO, under such conditions.

Thus it is not solely a matter of observing how much of the sky is occupied by CO 
emission, but also of assessing where the emission arises and  what it means beyond 
a simple scaling  to N(\HH).  The importance of accounting for \HH\ in
partially-molecular gas at low-moderate optical extinction 
is only reinforced by the recognition that the extinction is actually quite modest 
over large portions of large cloud complexes like Taurus \citep{GolHey+08} or
Chamaeleon \citep{Pla15Cham} and that conversion of CO brightness to N(\HH)
is fraught with difficulty \citep{GolHey+08} when isotopic fractionation and 
non-LTE  excitation are the rule.  

Moreover CO emission fails to represent substantial amounts of \HH\ in partially-molecular 
gas at the HI$\rightarrow$ \HH\ transition \citep{GreCas+05,WolHol+10}. 
When significant amounts of gas are  missed in a combined census of CO and HI emission 
-- the so-called dark neutral medium (DNM) \citep{Pla15Cham}  -- 
molecular gas is detectable in mm-wave absorption from 
\hcop, \cch\  and other small molecules \citep{LisGer+18} and the absence of detectable 
CO emission can be attributed to small CO column densities that are nonetheless
detectable in mm-wave absorption thanks to the power of modern mm-wave telescopes 
\citep{LisGer+19}. In Chamaeleon, the DNM missing in CO and/or HI emission in
the outskirts of the complex turns out to be  primarily molecular, even while 
the outlying medium is mostly but not overwhelmingly atomic.

In this work we present a somewhat eclectic (but ultimately coherent) 
assortment of new $\lambda$3mm emission-line observations of carbon monoxide 
and more strongly-polar molecules \hcop\ and CS toward and around previously-known 
\citep{LisPet12} and newly-identified positions having relatively intense CO 
emission at modest reddening, so-called CO hotspots where \WCO\ = 5-10 K-\kms.
Section 2 describes the 
circumstances of the new and existing observations that are discussed.
Section 3 is a source-by-source presentation of the data.  Results for 
the carbon monoxide isotopologues are gathered and modelled in Section 4.
Results for more strongly-polar species, chiefly the \hcop\ J=1-0 and CS J=2-1 
lines, are discussed in Section 5 where the CO hotspots around mm-wave
continuum background sources having \WCO\  = 5-10 K-\kms\ are interpreted as
regions of enhanced \HH\ and CO formation that are expected from models of 
turbulent gas at the HI$\rightarrow$\HH\ transition Section 6 is a summary
where we also discuss the relevance of the present results to observations
of the Milky Way and other galaxies.

\begin{table*}
\caption{Target, line of sight and map field properties}
{
\small
\begin{tabular}{lccccccccccccc}
\hline
Target & ra & dec       & l   &  b &  Map & \EBV$^1$ & N (H I)$^2$ & N(\HH)$^3$ & \fH2$^4$ & \WCO &  {X$_{\rm CO}$}$^5$ & X(CO)$^6$ \\
       & J2000  & J2000 &$^o$ & $^o$ & size & mag  & $10^{20}\pcc$ & $10^{20}\pcc$ & & K \kms  &  & \\
\hline
B0528+134 & 05:30:46.41 & 13:31:55.1 & 191.37 & -11.01 & 30\arcmin & 0.89 & 23.3 & 7.9 & 0.40 & 2.2 & 3.6 & 2.78\\
B0736+017 & 07:39:18.03 & 01:37:04.6 &  216.99 & 11.38 & 15\arcmin &0.13 & 8.0   & 3.3 & 0.45 & 0.8 & 4.1 & 2.42\\
B0954+658 & 09:58:47.24 & 65:33:54.7 & 145.75 & 43.13 & 30\arcmin & 0.12 & 4.7   & 4.8 & 0.67 & 1.6 & 3.0 & 3.33\\
B1954+513 & 19:55:42.69 & 51.31:48.5 & 85.30 & 11.76 & 30\arcmin  & 0.15 & 11.4  & 5.0 & 0.47 & 1.6 & 3.1 & 3.20\\
B2200+420 & 22:02:43.24 & 42:16:39.9 & 92.59 & -10.44 & 30\arcmin & 0.33 & 17.1  & 8.8 & 0.51 & 5.8 & 1.5& 6.59\\
B2251+158 & 22:53:57.71 & 16:08:53.4 & 86.11 & -38.18 & 30\arcmin & 0.10 & 7.2   & 1.2 & 0.25 & 0.8 & 1.8 & 6.66\\ 
L121 peak & 16:32:44.20 &-12:09:48.1 & 4.53  & 22.9   & 90\arcmin & 0.67 & 14.7  &     &      & 12.3 & &  \\ 
\hline
\end{tabular}
\\
$^1$from \cite{SchFin+98} \\
$^2$ N(H I) = $1.823 \times 10^{18}\pcc \int{\TB(H I)}dv$ with data from EBHIS or GASS III  \\
$^3$ N(\HH) = N(\hcop)/$3\times10^{-9}$ see Section 2.3 \\
$^4$ \fH2 = 2N(\HH)/(N(H I) + 2N(\HH)). $<$N(H)/\EBV$> = \Sigma$ N(H)/$\Sigma$\EBV\ $= 7.63 \times 10^{21}\pcc$ mag $^{-1}$  \\
$^5$ X$_{\rm CO}$} = N(\HH)/\WCO, units are $10^{20}\pcc$/(K-\kms). $<$X$_{\rm CO}> = \Sigma$ N(\HH)/$\Sigma$\WCO\ 
$= 2.42 \times 10^{20}\pcc$ (K-\kms)$^{-1}$ \\
$^6$ X(CO) =( \WCO$\times10^{15}\pcc)$/N(\HH))$\times10^6$,$<$X(CO)$>$ = $\Sigma$N(CO)/$\Sigma$N(\HH) = $4.1\times10^{-6}$
\\
\end{table*}

\begin{figure*}
\includegraphics[height=22cm]{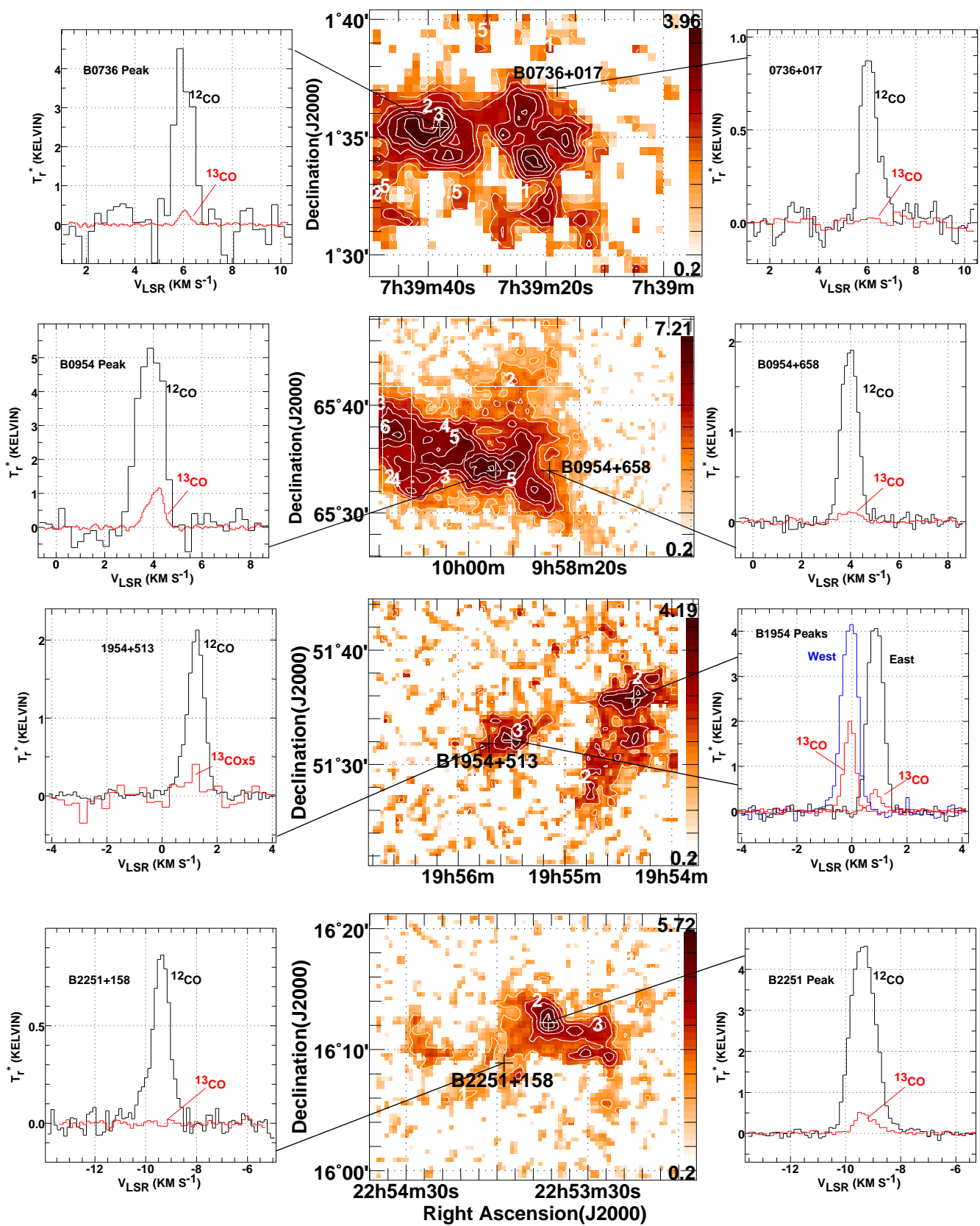}
\caption{\cotw\ and \coth\ emission around B0736, B0954, B1954 and B2251.  In the
center are maps of \WCO\ with the positions of the continuum source and one or
more  emission peaks indicated.  Spectra toward the continuum and the peak positions
are shown to either side.  The contours are at levels 1,2 .. 7 \Kkms\
and the color scale on the maps at center varies between 0.2 \Kkms\ and the
peak value as shown at the upper end.}
\end{figure*}

\section{Observations and modelling}

\subsection{New data}

The new data shown here were taken at the ARO Kitt Peak 12m telescope
during several observing runs in December or April during the period
2010-2012. Table 1 reprises some general target and sightline 
characteristics; coordinates, reddening, etc.  Table 2 shows positions and 
profile integrals for the individiual sightlines that furnish the observational 
material.

The ARO data are presented in terms of the customary brightness
scale \Tstar\ extrapolated outside the atmosphere and nominally
corrected  for  primary beam efficiency by an increase of a factor 
1/0.85.  As noted by \cite{LisPet+09} this scale is commensurate 
with the main beam brightness temperature scale of the Nanten telescope, as
empirically verified by comparing observations with both instruments
 over the L121 cloud near \zoph.  The LSR velocity scale uses the common
kinematic definition that is universally employed at radio telescopes.

The new observations mostly consist of pointed, position-switched  observations 
of species other than
\cotw\ toward peaks found in maps of \cotw\ emission around the positions of 
compact extragalactic continuum sources used to study mm-wave absorption.
The \cotw\ maps  used as finding charts  were discussed in \cite{LisPet12} 
 and the new observations should be understood in context with reference 
to their Figures showing the placement of the background targets maps of 
galactic reddening, noting their separation from regions of high extinction.
However, we also observed
at positions of \cotw\ peaks toward portions of the L121 cloud that we identified 
in new ARO Kitt Peak 12m maps of \cotw\ emission, and we made 
an on-the-fly map of \coth\ emission in the field around B0528-134.  Nearby 
off-source positions for position-switching were already known from the 
extensive mapping we had done.  

Observations toward all sources are discussed individually in the ensuing text 
but the overall goal of understanding the sytematics of CO emission from 
diffuse molecular gas is really only approachable when the full ensemble of 
observations is examined.
 
\begin{figure*}
 \includegraphics[height=16.3cm]{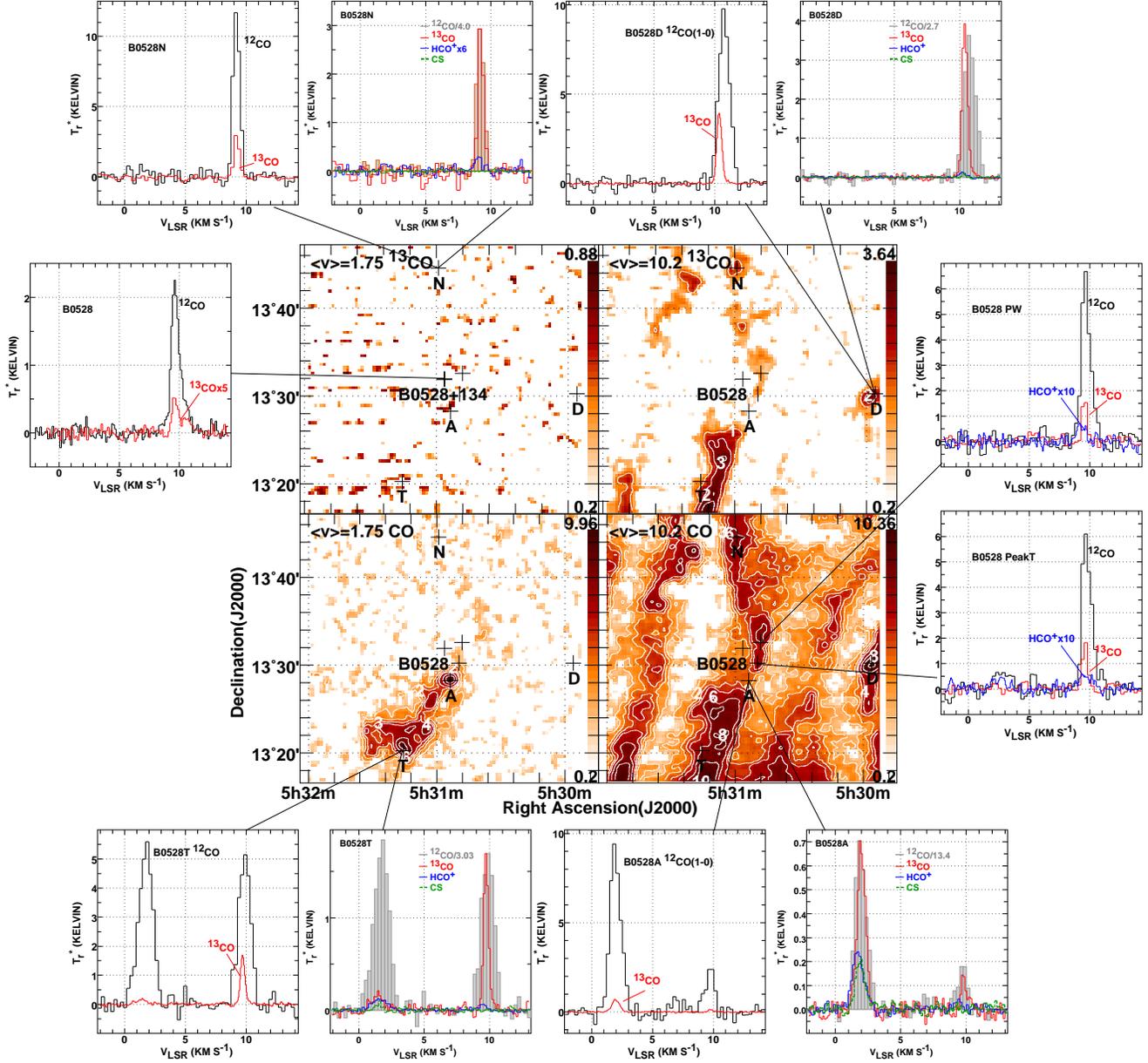}
  \caption{Molecular emission around B0528+134.  At center are maps of
\coth\ in the upper row and \cotw\ at bottom, with the blue shifted gas at
left and the red-shifted emission at right. The continuum and other selected 
locations are marked, with lines directed to the spectra at those positions.  
In some cases the same position is represented by two plots of spectra.  Contours
are at levels 1,2, 3, 4, 6, 8 and 10 \Kkms.
}
\end{figure*}

\begin{figure*}
 \includegraphics[height=11.7cm]{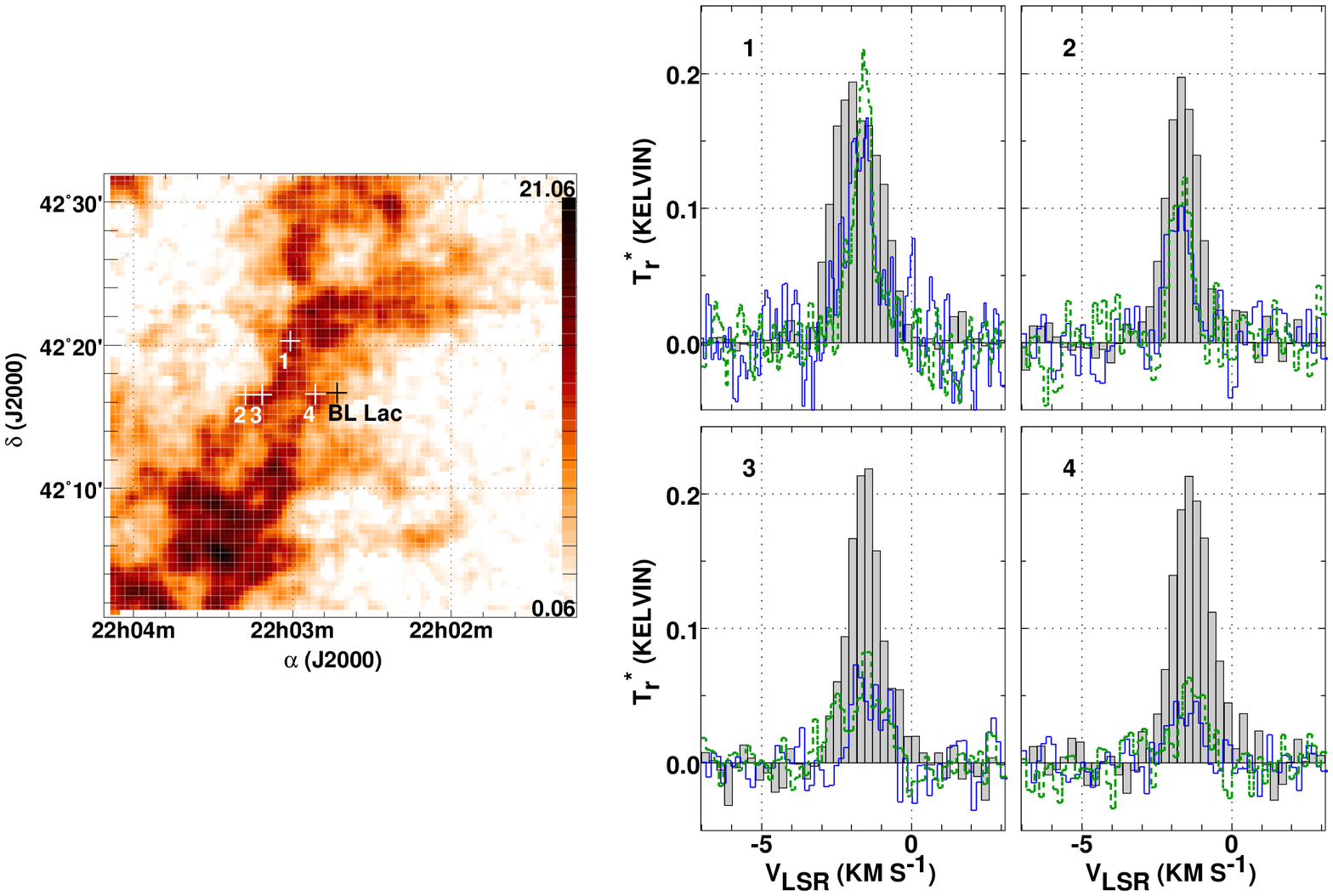}
  \caption{Molecular emission around \bll.  At left is a map of \WCO,
on which are marked the locations of \bll\ and four positions whose
CO, \hcop, and CS spectra are shown at right.  The CO profiles at right,
shown gray and shaded, have been scaled downward by a factor 30. \hcop\ is
shown as a blue histogram with full lines, CS is shown dashed and green.}
\end{figure*}

Most of the new CO observations are of \coth, as the CO hotspots had
previously only been observed in \cotw.  But we also searched for \coei\
emission in some directions and we made extensive
observations of the J=1-0 line of \hcop\ and more limited observations of
CS J=2-1.  The 1-0 lines of HCN, HNC, and C$_2$H were observed toward
Peak 1 near \bll\ and are shown for the sake of completeness in Appendix B 
where their profile integrals are given.  The \WHCN/\WCO\ ratio at Peak 1 
around \bll, $0.017 \pm 0.001$, is discussed in Section 6.7.

All of the new pointed
observations were observed at higher spectral resolution and smoothed to 49 kHz
resolution or 0.13 \kms\ at 115.3 GHz.  The \coth\ OTF mapping around B0528
had a resolution of 100 kHz or 0.27 \kms.  In the comparisons below, some 
CO peaks were only observed in \cotw\ during the earlier OTF mapping 
and these profiles have a resolution 100 kHz (0.26 \kms\ at 115.3 GHz). 

\subsection{Existing data}

We show emission profiles from the CO mapping around continuum sources
from which many of the peaks observed here were first identified \citep{LisPet12}
and we show pointed observations taken with higher spectral resolution
toward the continuum sources themselves. We discuss CO line profile 
integrals toward a wider sample of continuum background sources observed by 
\cite{LisLuc98} (labelled PdBI or PdBI sightlines) and from the CO emission survey 
toward common UV absorption targets of \cite{Lis08} and labelled 
``Survey 2008''.  Some CO emission data closer to \zoph\ were discussed by 
\cite{Lis97} and these are labelled \zoph $\pm30$\arcmin.  We also discuss the 
``Mask 1'' subset of CO emission brightnesses at low extinction in Taurus 
observed by \cite{GolHey+08}.

In Section 6 we derive molecular hydrogen number and colum densities by comparing
emission and absorption profiles of the \hcop\ J=1-0 line, mostly taken from the
work of \cite{LucLis96} but augmented in a few cases with newer NOEMA absorption 
profiles of higher sensitivity and spectral resolution.

\subsection{Modeling}

In what follows we  compare the observations with recently-published models 
of CO formation, excitation and fractionation \citep{Lis17Frac}.  The models 
self-consistently calculate the equilibrium abundances of \HH\ \citep{Lis15HD} 
and CO \citep{Lis07CO} and their isotopologues in uniformly illuminated,
uniform density media having spherical geometry, in the presence of 
photodissociation, self- and mutual shielding, and exchange of carbon 
isotopes by the reaction of C\p\ and CO.  All aspects of the illumination 
(optical/UV, cosmic-ray, X-ray) are scaled by a parameter \G0\  = 1/3, 1/2 and 1.
The CO chemistry is modeled under 
the simple assumption that the observed quantity of 
\hcop, N(\hcop)/N(\HH) $= 3\times 10^{-9}$ (\cite{LisPet+10} and \cite{LisGer16}) 
recombines to form CO at the local kinetic temperature that is self-consistently
computed in the models \citep{WolHol+95,WolMcK+03}. It is actually possible to 
check the usefulness of this approach given the observed \hcop\ emission as 
described in Section 5.1.  This is because the formation of CO from the 
recombination of \hcop\ and the rotational excitation of \hcop\ both arise 
from the interaction of \hcop\ with free electrons. 

The  isotopologic abundance ratios \hcop/H$^{13}$CO\p/HC$^{18}$O\p\
in the model are taken as 1:1/62:1/520 so that the CO isotopolgues are 
formed in the same proportion as the inherent isotopic abundance ratios 
in the gas.  It is not possible to fractionate \hcop\ through its chemical 
isotopic exchange reactions with CO \citep{RouLoi+15} in diffuse gas where 
the electron abundance is high, most of the carbon is in C\p\ and the 
abundance of CO is small: This is the case because, unlike in dark
clouds,  \hcop\ recombines 
with electrons thousands of times more rapidly than it reacts with CO. 
The CO isotopologues are fractionated after formation through the 
processes of selective photodissociation and carbon isotope exchange 
at the local kinetic temperature.

\section{Fields observed}

The observed sample of CO hotspots is comprised of two parts; i) observations 
of gas within 
5-30\arcmin\ of the positions of compact extragalactic continuum sources 
used to study molecular absorption at radio wavelengths, and ii) observations 
over a region of extent 90\arcmin\ toward the translucent cloud L121 
southwest of the star \zoph.  Except for the continuum source B0528-134
that is observed through the Eridanus Loop, the gas seen around the 
continuum background sources is well removed from regions showing 
\AV\ $ \ga$ 1 mag. As shown in Figure 1 of \cite{LisPet12}, 
B0954+658 (\EBV\ = 0.12 mag) and B2251+158 (\EBV\ = 0.10 mag) are within 
a few degrees of MBM 23 and MBM 55 \citep{MagBli+85} having \EBV\ = 0.13 
and 0.29 mag, respectively. B1954+513 (\EBV\ = 0.15 mag) and B0736+017 
(\EBV\ = 0.13 mag) are more isolated. The material studied here 
samples the diffuse, occasionally partially-molecular ISM that is 
broadly distributed outside dark/molecular cloud complexes.

By contrast, L121 in Ophiuchus is only 1\degr-2\degr\
removed from a very extended optically-opaque filament L204
whose CO emission is redshifted by some 4 \kms, as shown in 
Fig. 1 of \cite{LisPet+09}.  The fact that L121 and L204 had
such comparable CO brightnesses was a strong motivating factor
for this work.

\subsection{Four fields around compact mm-wave continuum sources}

\begin{figure*}
 \includegraphics[height=10.5cm]{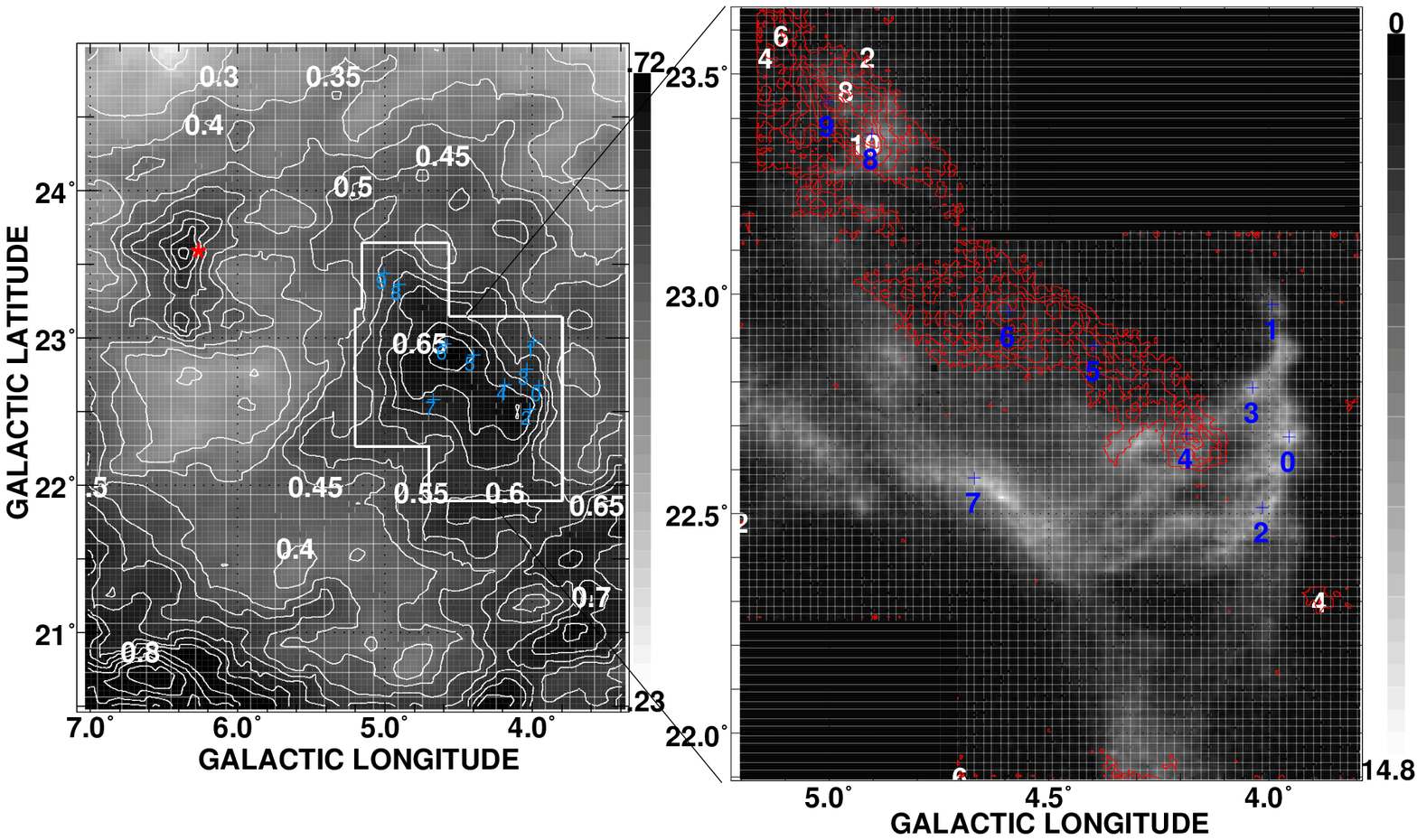}  
\caption{Reddening and CO emission in the vicinity of \zoph, L121 and L204
\citep{LisPet+09}. Left side panel: A map of \EBV\ \cite{SchFin+98} with 
contours in steps of 0.05 mag and a gray scale running from 0.23 to 0.72 mag 
as indicated on the adjacent bar scale. The position of the star \zoph\ is 
marked by a red asterisk and the portion of L121 mapped in \cotw\ emission 
at the ARO 12m at 1\arcmin\ spatial resolution is outlined. Indicated 
in blue in both panels are 10 labelled locations Z0..Z9 ordered in galactic 
longitude where additional pointed observations with higher spectral 
resolution were taken (see Figure 5 and Table A1). A small part of the  
filamentary cloud L204 is visible as the high extinction region at lower left.
Right side panel:  ARO \cotw\ maps of CO emission from L121 at 1\arcmin\ 
resolution on an expanded spatial scale.  This panel is a composite of a 
gray scale representation of integrated emission at v $<$ -0.7 \kms\ 
with peak \WCO\ = 14.8 \Kkms, and contours in red at levels 2, 4, .. 14 
\Kkms\ representing the integrated emission at v $> -0.7$ \kms.} 
\end{figure*}

Figure 1 shows results for CO and \coth\ around the positions of four
extragalactic continuum sources.  In the center column are cut-outs of the 
larger CO maps shown in \cite{LisPet12} with peak positions and the
location of the continuum marked by crosses. The peak integrated
brightness of the \cotw\ emission is shown at the top of the color
bar scale at the right side of each map: these vary from 4-7 K-\kms. 
To either side, profiles taken toward the continuum or at the nearby peaks 
are shown, with connecting lines drawn
between the positions of CO peaks and the boxes containing the profiles 
observed there.  As noted in the earlier work, CO lines have typical 
brightness 1-2 K toward the continuum, and CO peaks of remarkably similar
brightness 4-5 K nearby.  \coth\ is detected at most of the positions
and is remarkably bright toward the western, blue-shifted CO peak 
near B1954+513.  This was by chance the very first position observed
in \coth\ in the course of this work.  However similar the CO brightnesses
of the peaks may be around B1954+513, the \coth\ brightness varies by more 
than a factor 5.

\subsection{The field around B0528+134}

A more extensive set of observations around B0528+134 is shown in Figure 2.
In the center are four map panels representing \coth\ emission at top
and \cotw\ below, each species sampled over two velocity ranges 
corresponding to the blue-shifted gas around 1.75 \kms\ at left and 
the redshifted component at 9-12 \kms\ at right.
Connecting lines are drawn between map locations and the boxes containing
the spectra there, and the same position is sometimes represented by
two boxes with different contents and/or scaling of the brightness.

The line of sight to B0528+134 passes near the Eridanus Loop and is
not characterized by especially small extinction; \EBV\ = 0.89 mag in 
Table 1.  However, the \hcop\ column density measured in absorption is 
smaller than that seen toward \bll\ at \EBV\ = 0.32 mag, indicating
locally diffuse conditions.  The CO emission pattern around B0528+134
is characterized by very regular striations of a component at
9-12 \kms\ having high contrast and very high CO peak brightnesses 
typically near 10 K, shown over a wider field by \cite{LisPet12}.  
This may be the second example of a 
phenomenon originally seen in Orion \citep{BerMar+10} and ascribed
to `` `waves' at the surface of the Orion molecular cloud, near where 
massive stars are forming.'' Alternatively, striations in molecular gas
are discussed by \citep{HeyGol+16} and \citep{TriFed+18} and ascribed
to some combination of instability, MHD waves and alignment with the magnetic
field.  

\begin{figure*}
 \includegraphics[height=13cm]{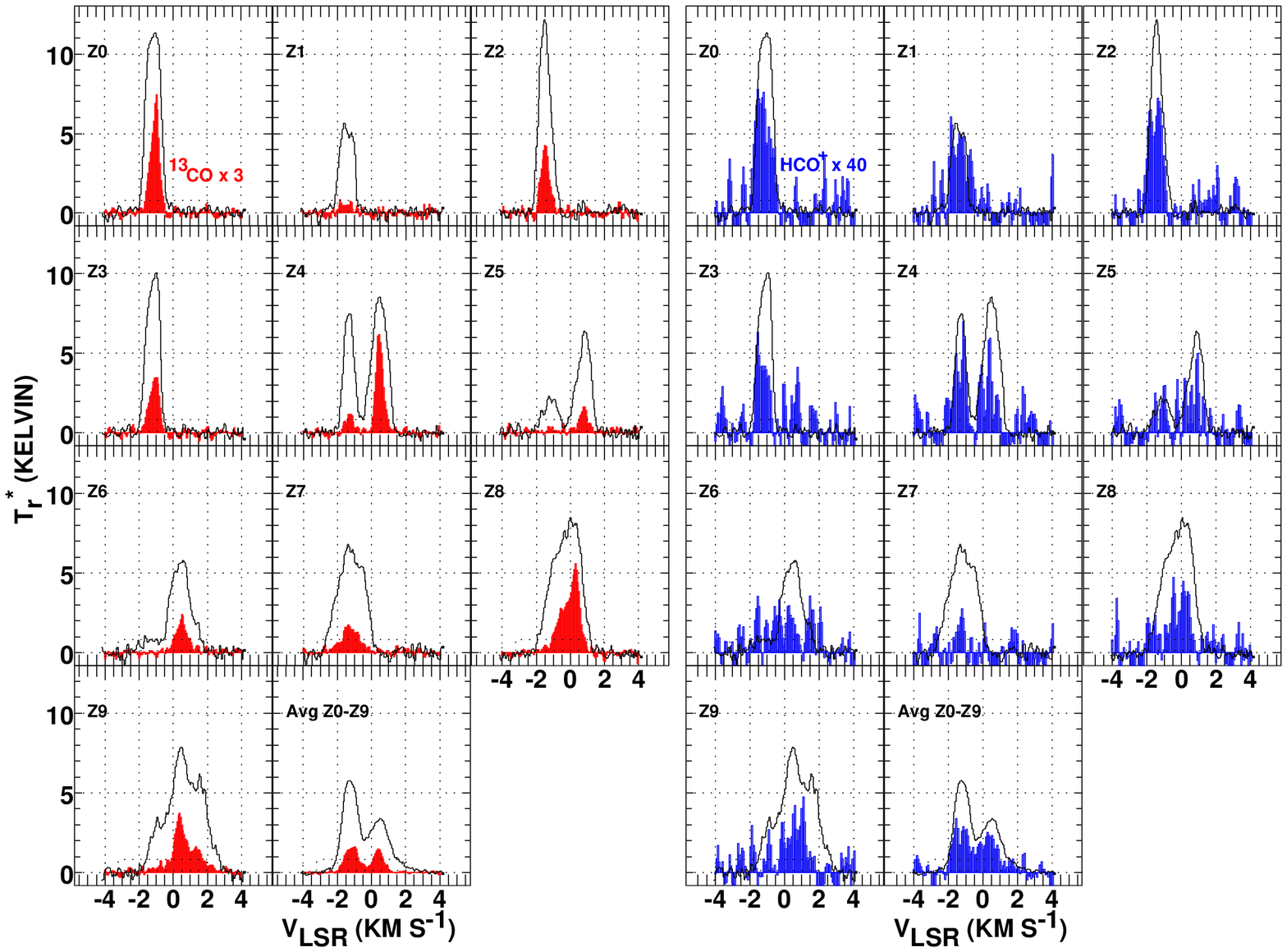}  
\caption{\cotw, \coth, and \hcop\ emission from 10 pointings in L121 and
their unweighted average at bottom right in each panel.
At left \coth\ emission scaled upward by a factor 3 is shown in
red overlaid on CO, while at right \hcop\ shown in blue and scaled upward 
by a factor 40 is overlaid on CO.} 
\end{figure*}

As shown in the CO maps in the lower row at the center in Figure 2,  
the blue-shifted component overlaps a limited portion of
the red-shifted emission to the southeast of the continnum and is absent
elsewhere. This similarity of morphology suggests a physical connection
but the conditions in the red and blue-shifted gas are dissimilar.
 Comparing the upper and lower maps in either column, one sees 
that the blue-sshifted gas in the left-hand column is much fainter in \coth.
Observations at the position B0528T shown in boxes at the lower left
corner of Figure 2 illustrate the complexity of the variation in the
line brightneses.  At a position where both the red and blue components
are equally bright in \cotw\ the red peak is much brighter in \coth.
To the blue where \coth\ is weaker, \coth, \hcop\ and CS are all about
equally bright.  To the red, \coth\ is an order of magnitude brighter
than \hcop\ and CS is not detected.  Variations in the \hcop/CS brightness
ratio are indicative of large differences in their relative abundances
as discussed by \cite{Lis12} and in more detail in Section 5.2.

\subsection{The field around \bll}

Figure 3 shows profiles at four positions near \bll, as marked on
a cutout of the \cotw\ map shown in \cite{LisPet12}.  Profiles
of \cotw\ scaled downward by a factor of 30 are shown as grey histograms
with profiles of \hcop\ J=1-0 in red and CS J=2-1 in blue and dashed.
Unlike most of the other directions observed here, ie the positions 
sampled around B0528-134 in Figure 2 (see Section 5.2), \hcop\ 
and CS emission have comparable brightness at all four locations.
\coth\ observations at these positions were thwarted when a snowstorm
bore down on the mountain on the day of the Newtown School shootings
in December 2012 during the final observing run of this program.  The
peak integrated brightness of the \cotw\ emission is 21 K-\kms, which
seems remarkable given that the peak redddening is \EBV\ = 0.36 mag 
in this region \citep{LisPet12}.

\subsection{L121}

\begin{figure*}
\includegraphics[height=9cm]{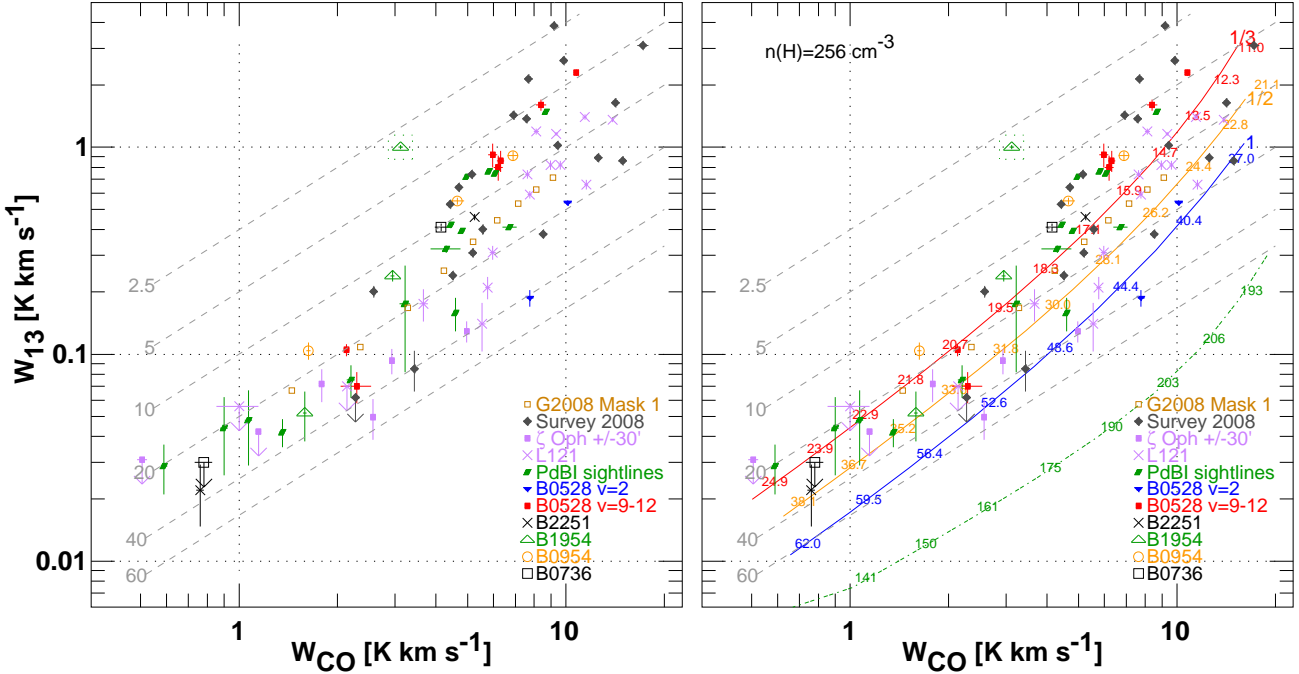}
\caption{\coth\ and \cotw\ emission for various datasets.
Fiducial values of the \WCO/\Wth\ ratio are shown as grey dashed lines.
Right: with model results superposed for n(H) = 256 $\pccc$ and three 
scaled strengths of the illuminating radiation
field, at full, one-half and one-third strength (\G0\  = 1, 1/2, and 1/3 respectively).
Results neglecting isotope exchange for \G0\ =1 are shown
as a green dash-dotted line.  The numbers along the model curves are the
N(CO)/N(\coth) column density in the model, for comparison with the intensity ratio that
is plotted.  
In general the brightness and column density ratios track each other even
when they are much smaller than the inherent isotopic abundance ratio of C/$^{13}$C = 62.}
\end{figure*}

Figure 4 gives an overview of the observing done in the L121 region
near \zoph\ \citep{LisPet+09}.  At left is a map of reddening \EBV\
using the data of \cite{SchFin+98} with the star marked by a red
asterisk to the northeast.  The minimum reddening 0.23
mag must largely  be associated with foreground and background material:
The material associated with L121 contributes about 0.65-0.23 = 0.4 mag
of reddening, or some 1.2 magnitudes of visual extinction in a region where 
the peak reddening \EBV\ =  0.65 mag indicates 2 magnitudes
 of visual extinction.

CO was mapped over the region outlined at left in Figure 4 by stitching 
together numerous OTF tiles of size approximately 30\arcmin\ x 30\arcmin.
The grey scale 
at right with a peak of \WCO\ = 15 K-\kms\ represents the integrated
CO profile of  the blue-shifted CO emission at v $\le -0.7$ \kms: 
the contours in red are for the red-shifted emission at v $> -0.7$ \kms\
that is rather distinct from the more blue-shifted emission.  Both
components are observed in optical absorption and radiofrequency emission 
toward the star \zoph\ itself. Numbers in white in
the right-side frame in Figure 4 are the integrated brightness
of the red-shifted emission which peaks just above 10 K-\kms\ to the north.

Indicated at left and right in Figure 4 are ten locations labeled 0..9
in order of increasing galactic longitude at which pointed observations
of CO, \coth\ and \hcop\ were taken, as shown in part in Figure 5.  Positions
0 and 9 were observed in \coei\ and a few positions were also observed in CS.
CO, \coth\ and \hcop\ profiles toward the 10 enumerated positions
and their unweighted mean are shown in Figure 5.  At left, \coth\ profiles 
scaled upward by a factor three are overlaid on \cotw.  At right, \hcop\ 
profiles scaled upward by a factor 40 are overlaid on \cotw.  As with 
the other fields, CO components with similar \WCO\ have widely varying 
\coth\ brightness.

\section{Results for CO isotopologues}

\subsection{C$^{13}$O and $^{12}$CO}
 
Figure 6 shows a summary of the new and existing observations of CO and \coth\
in diffuse molecular gas:  the new data discussed here have been added to the 
comparable Figure 4 in \cite{Lis17Frac}. The brightest \cotw\ lines in
diffuse/translucent gas generally have \WCO\ $\approx$ 10 K-\kms\ corresponding to 
N(CO) $\approx 10^{16}\pcc$.  Exceptions occur 
in the larger field around \bll\ (Figure 3),
toward L121 when the CO line is broad and the red and blue shifted components cannot 
be separated (Table 2 and Figure 5), and for a few sightlines toward stars used as 
UV absorption line targets. Many of the smallest ratios \WCO/\Wth\ $< 5$ 
and the largest \WCO\ values arise along sightlines toward UV absorption line 
targets, in which case CO emission likely originates from dense placental material 
situated behind the stellar targets: stars seen behind gas emitting very bright 
CO lines would likely have been too heavily extincted to serve as good UV
absorption line targets.   

\begin{figure}
\includegraphics[height=8.3cm]{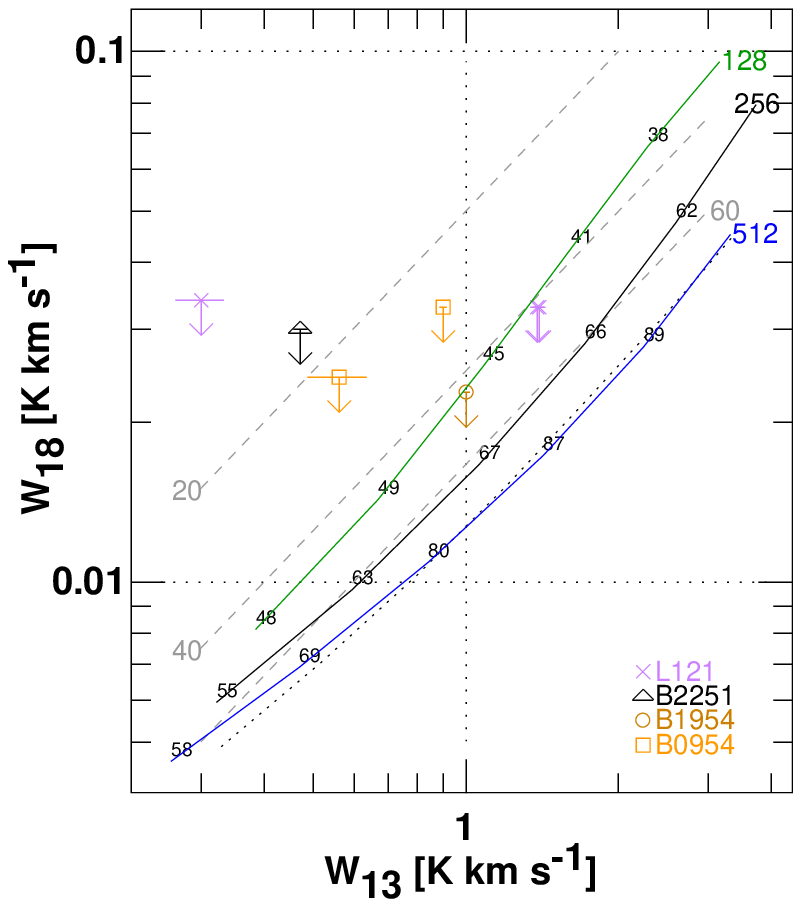}
\caption{Observed and calculated \coei\ and \coth\ emission profile integrals.  
Fiducial values of the ratio \Wth/W$_{18}$\ = 20, 40 and 60 are shown as grey dashed 
lines.  Also shown are  calculated brightnesses for n(H) = 128 (green), 256 (black) and 
512 (blue) $\pccc$ at full strength and as a dotted black line for n(H) = 256 $\pccc$ 
with the radiation field at half-strength, showing that the brightness ratio scales as 
\G0\ /n(H).   Numbers along the curves indicate the intrinsic N(\coth)/N(\coei)
ratio.  The intrinsic $^{13}$C$/^{18}$O ratio in the gas is 520/62 = 8.4.}
\end{figure}

With the notable exception of one datapoint with very strong \coth\ emission 
near B1954+513 (Figure 1), the diverse observational samples provide a 
consistent overall picture whereby 
\Wth/\WCO\ $\approx 1/40-1/20$ at \WCO\ $\la 4$ K-\kms, increasing so that 
\Wth/\WCO\ $\approx 1/20-1/5$ at \WCO $\ga 4$ K-\kms.

To understand the observations we show at right in Figure 6 the same data 
with model results for gas of total density n(H) = 256 $\pccc$ and three 
values of the scaled strength of the incident radiation field
\G0\ =1, 1/2 and 1/3: the underlying assumptions and methods of the models are 
discussed in Section 2.3. Along each model curve there are numbers 
representing the ratio of column densities N(\cotw)/N(\coth) in the model.  
These can be compared with the intensity ratios in the plot to assess the 
degree to which the CO column densities and emission brightnesses track each
other, as stressed in \cite{Lis17Frac}.  To aid in this comparison, fiducial 
lines are shown at fixed intensity ratios 20, 40 and 60. 

In the righthand panel in Figure 6
the lowest-lying model curve shows results when the carbon isotope exchange is 
ignored:  Clearly, the success of the models depends most nearly on this process, 
and very nearly all of the observed \coth\ is formed from the reaction  of 
$^{13}$C\p\ with \cotw, not from the recombination of H$^{13}$CO\p\ with 
 electrons. A comparable situation arises for HD that is almost entirely 
formed {\it in situ} in the gas through gas-phase exchange reactions of
deuterium and hydrogen \citep{Lis15HD} although, in that case, HD
is largely self-shielding \citep{LePRou+02}.

The effects of changing the density 
and radiation field are complex in models where the heating is largely 
through the photoelectric effect on small grains, from photons of about
the same energy as those that ionize carbon and photodissociate \HH\ and 
CO.  Slower photodissociation and lower kinetic temperature each encourage 
more effective carbon isotope exchange while lower temperatures cause
faster recombination of \hcop\ to CO, weakening the fractionation
while allowing a given N(CO) to form at lower n(H), N(H) and N(\HH).
At a given \WCO\ the \Wth/\WCO\ integrated brightness ratio increases with 
higher density and weaker illumination, with a progressively stronger
density dependence at amaller values of \G0\ .
The highest observed  values of \Wth/\WCO\ require some combination of 
higher density and weaker radiation and the effect of dimming the radiation 
is stronger at the higher densities. 

For \WCO\ $\la$ 4 K-\kms\ where \Wth/\WCO\ $\approx 1/40-1/20$ the
data can generally be explained at any number density 
n(H) $\ga 128\pccc$, but even the lowest of these 
densities is already large enough that the molecular fraction \mfH2\ 
= 2N(\HH)/N(H) $\ga$ 0.7 according to Figure 6 of \cite{Lis17Frac}. 

The isotopologic intensity ratio 
\WCO/\Wth\ tracks the column density ratio N(CO)/N(\coth) even when both 
ratios are much less than 62 and  the proportionality between brightness and column
density persists well beyond the domain where $\tau > 1$ as long as the
excitation is strongly sub-thermal \citep{GolKwa74}. The difference between 
\WCO/\Wth\ and N(CO)/N(\coth) is about 20\% at \WCO\ = 5 K-\kms\ or
40\% at \WCO\ = 10 K-\kms, with slightly larger differences at smaller
\G0\ . The point is that when the isotopologic brightness ratio significantly
departs from 62, the actual abundance ratio is much closer to the observed 
brightness ratio than to the intrinsic isotopic abundance ratio.
N(\cotw) cannot be reliably estimated from comparing the \cotw\ and \coth\ 
brightnesses under the common assumption that both isotopologues are present in 
the intrinsic ratio, and the wide variations in N(CO)/N(\HH) in any
case mean that N(CO) cannot be used to infer N(\HH) in any case.
This point was stressed by \cite{SzuGlo+16} who concluded that it
was more accurate to derive N(\HH) by scaling \WCO\ with a 
CO-\HH\ conversion factor.  If desired, the carbon monoxide
column densities can  be estimated as N(CO) $= 10^{15}\pcc$ \WCO\
and similarly for other isotopologues.

Two aspects of the data in Figure 6 deserve comment insofar as they relate 
to a comparison with UV absorption line data as discussed by \citep{Lis17Frac}.  
The nearly
constant ratio N(CO)/N(\coth) $\approx 65$ seen in UV absorption 
at N(CO) $< 10^{16}\pcc$ is higher than that inferred from mm-wave 
emission brightness ratios and corresponds most closely to the curves for 
\G0\ =1: The gas observed in optical absorption is mostly unfractionated, 
from which we infer that it is subject to a stronger radiation field 
and is warmer and less dense than that observed in mm-wave emission.
Conversely, the CO emission data taken in directions toward the same
early-type stars that are observed in UV absorption \citep{Lis08} 
require a diminished radiation field at the density n(H) = 256 $\pccc$ 
that was used in Figure 6.  The counter-intuitive inference of a weaker
radiation field is consistent with the notion that strong CO emission
toward the absorption line stars arises in dense fully-molecular gas 
behind the star because these stars would not
have been suitable UV absorption targets with the opposite geometry.  
\zoph\ is a case where little or none of the molecular emission observed 
in its direction arises behind the star, but the emission is weak, 
\WCO\ $\approx$ 1 K-\kms.

\begin{figure*}
\includegraphics[height=8.2cm]{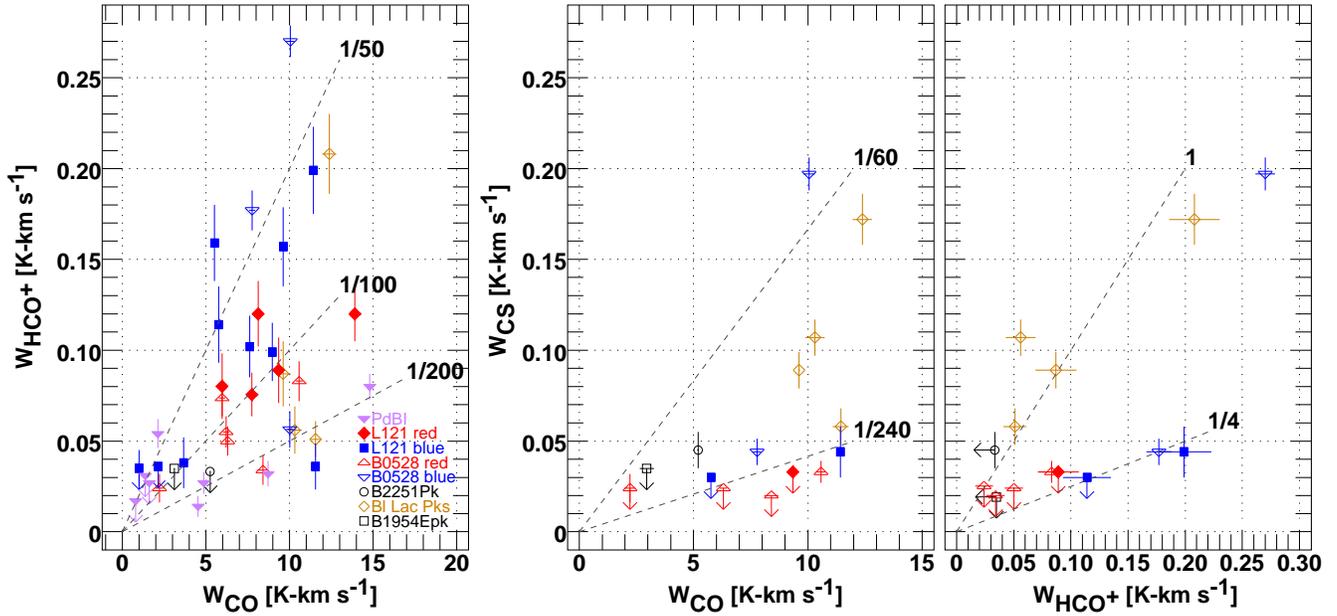}
\caption{Observed \cotw, \hcop, and CS profile integrals. Fiducial values 
are shown as grey dashed lines.}
\end{figure*}
 
\begin{figure}
\includegraphics[height=8.2cm]{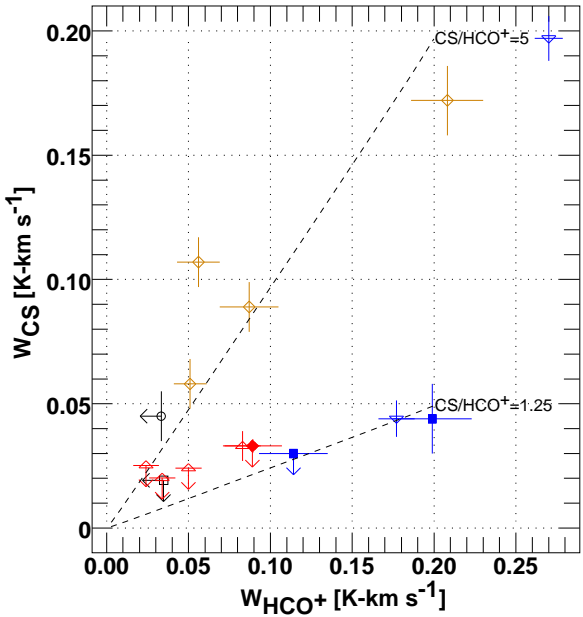}
\caption{As in the right-most panel of Figure 8, but with model results 
for two values of the abundance ratio, CS/\hcop\ = 5 and 5/4.  Calculated
results are not perceptibly different for different density n(H) and 
illumination \G0\ .}
\end{figure}
 
\begin{figure}
\includegraphics[height=7.5cm]{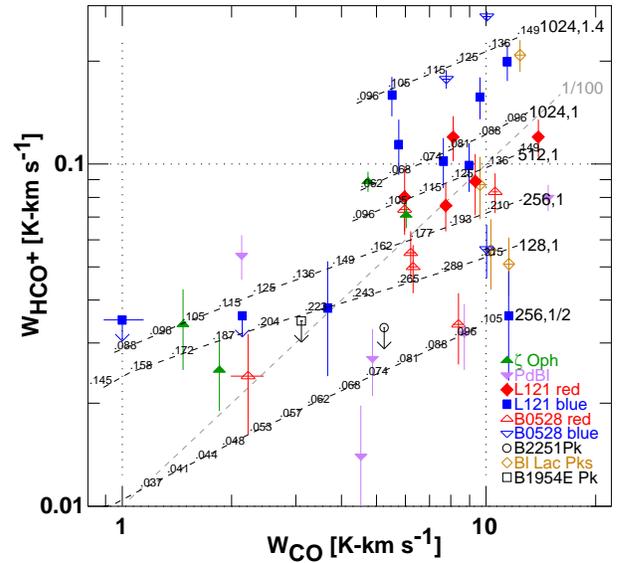}
\caption{Observed \cotw\ and \hcop\ line profile integrals on a logarithmic 
scale with model results. Labels at the right hand side of each model
curve represent n(H) in units of $\pccc$ and the radiation field scaling
parameter \G0\ . Numbers along the model curves are the reddening \EBV.
A fiducial line with \Whcop/\WCO\ = 0.01 is shown grayed and dashed.}
\end{figure}

\subsection{C$^{18}$O}

Few measurements of absorption by \coei\ are available in the UV, but two 
sightlines with N(CO) $= 2\times10^{15}\pcc$ and $= 2.5\times10^{16}\pcc$
show \coei\ depletions of 3 and 6, respectively, with respect to the ratio
N(CO)/N(\coei) = 520 (see Figure 1 of \citep{Lis17Frac}).  The mm-wave 
measurements presumably sample gas with weaker radiation fields and 
weaker  selective photodissocation of \coei.  


Figure 7 shows the limited amount of meaningful data we have been able to gather 
on the integrated intensity of \coei, W$_{18}$, compared with \Wth. Also shown 
in Figure 7 are calculated brightnesses for n(H) = 128, 256 and 512 $\pccc$ with 
the radiation field at full strength and at n(H) = 256 $\pccc$ with \G0\  = 1/2, 
Given along the curves showing model results are the intrinisic N(\coth)/N(\coei) 
ratios: these can be compared with the model brightness ratios to
see that the observed brightness ratios can straightforwardly be 
interpreted as column density ratios. The model brightnesses are weakly 
dependent on \G0\  and n(H) in opposite senses, falling with \G0\ , and 
the model brightness ratio scales approximately as \G0\ /n(H) (the curves coincide 
for n(H)=512 $\pccc$, \G0\ =1 and n(H)=256 $\pccc$, \G0\ =1/2). 

\coei\ emission is very weak in diffuse molecular gas, with $3\sigma$ 
observational upper limits \Wth/W$_{18} > 25-40$ in several cases as against 
an intrinsic isotopic abundance ratio $^{13}$C/$^{18}$O = 520/62 = 8.4. If 
N(\coth) is enhanced by a factor g, the implied \coei\ depletion factors 
are (25-40)/8.4/g = (3-5)/g. The limited range of relatively large \Wth\ 
values sampled in Figure 7 corresponds to high values g $\ga 2-4$ 
(see Figure 6) so the mm-wave measurements disappointingly do not constrain 
the \coei\ depletion at the levels observed in UV absorption spectra.


\section{Emission from strongly-polar species HCO\p\ and CS}

Figure 8 summarizes the \hcop\ and CS J=2-1 observations with respect to each
other and with respect to CO.  The panel plotting W$_\hcop$ vs. \WCO\ at left 
shows  that \hcop\ brightens considerably for \WCO\ = 
6 - 10 K-\kms\ but W$_\hcop$/\WCO\ = 1/200 - 1/50 quite generally.
Such ratios are characteristic of gas in the disk of the inner Milky Way,  
outside the galactic center \citep{Lis95,HelBli97}, and other galaxies 
\citep{JimBig+19} as discussed in Section 6.8.

CS J=2-1 emission brightens abruptly at \WCO\ $\ga 10$ K-\kms\ in the middle 
panel of Figure 8, from data in the blue component around B0528 (see Figure 2)
and around \bll\ (Figure 3).  The rightmost panel in Figure 8 shows that the 
W$_{\rm CS}$/\Whcop\ ratio is bimodal. The same datapoints toward \bll\ and 
in the blue component around B0528 having stronger CS emission at \WCO\ = 
10-12 K-\kms\ in the middle panel have three-four times larger 
W$_{\rm CS}$/\Whcop\ ratios than the rest of the data, over a wide range of 
\Whcop.

The \hcop\ J=1-0 and CS J=2-1 line brightnesses  are small enough that
both are in the weak excitation limit of \cite{LisPet16} and the line 
profile integral may be expressed as a  line of sight integral
through the emitting medium

 $$\int \TB d{\rm v} = (\lambda^2/8\times 10^5 \pi)(hc/k)                                                                                      
  \int n(H)  \gamma_u n(Y) dL   \eqno(1) $$

where Y stands for \hcop\ or CS and $\gamma_u$ is an excitation rate 
per H-nucleus determined by the local temperature and ionization fraction.  
The emergent line brightness is roughly proportional 
to the product of the mean number density of hydrogen and the column
density of the emitting molecule, not its critical density.
Details of the cloud structure and 
molecular abundance distribution are washed out by the line of sight 
integration and the emergent line brightnesses are not explictly
dependent  on the optical depth.
 
\subsection{\WCS/\Whcop\ and N(CS)/N(HCO\p) are bimodal}

The bimodality of the \WCS/\Whcop\ brightness ratio in the right-most 
panel of Figure 8, with values approximately 1 and 1/4, reflects a real 
bimodality of the N(CS)/N(\hcop) abundance ratio: there is no effect
associated with the excitation that can cause such a difference given
that both species must coexist in the regions that have higher \HH\ fractions.
Shown in Figure 9 are 
results for the model brightnesses of CS J=2-1 and 
\hcop\ under the assumption that n(CS)/n(\hcop) = 5 and n(CS)/n(\hcop) = 5/4
everywhere in the models (recall that n(\hcop)/n(\HH) $ = 3\times 10^{-9}$ is
fixed while n(H) and n(\HH) vary) and the model results for the intensity ratio 
are insensitive to n(H).

As discussed in Appendix C, bimodality of the N(CS)/N(\hcop) ratio was not 
observed in our survey of mm-wave \hcop\ and CS absorption \citep{LucLis02}, 
because the sightlines directly toward continuum sources sample material with 
CO emission well below the \WCO\ values  where bimodality is evident in Figure 
8.  The previously-observed column density ratios N(CS)/N(\hcop) derived in 
absorption have mean $<$N(CS)$>$/$<$N(\hcop)$>$ = 1.27 
and $<$N(CS)/N(\hcop)$>$ = 1.66$\pm$1.32 and are in overall 
agreement with the column density ratios derived now in emission alone.

Bistability is a recognized phenomenon in chemical networks for dense gas
 \citep{LeBPin+93,LeeRou+98,ChaMar03,BogSte06,WakHer+06,DufCha19} and the abundance of 
CS is one of the earliest recognized markers \citep{LeBPin+93}. \cite{GerFal+97}
found a column density of CS in a core in Polaris at \EBV\ $\approx 0.6$ mag 
that was 20-40 times higher than in Taurus at \EBV\ $\ga 1$ mag: the
N(CS)/N(\hcop) ratio was about a factor two higher in Polaris even as the 
brightness ratio \WCS/\Whcop\ was smaller by a factor three, 
being 2 in Taurus and 0.7 in Polaris.  The kind of
bistability that is discussed for dense gas chemical networks with a contrast
between high and low ionization phases driving changes in the abundance of
{H$_3$}$^+$ is not obviously relevant in diffuse molecular gas where CO is a 
minority constituent, the ionization fraction is high and electron
recombination of {H$_3$}$^+$ is rapid.

\subsection{HCO\p\ and CO}

On its own terms the \hcop\ line brightness is subject to the same considerations 
of the weak excitation limit as CS, but the special role of \hcop\ as the 
progenitor of CO and its fixed abundance with respect to \HH\ mean that the
comparison of \WCO\ and \Whcop\ has broader implications for the CO formation
chemistry. 
 
As a first approach to interpret the \hcop\ data with respect to CO, recall that 
the rotational excitation of \hcop\ is dominated by collisions with electrons in 
diffuse molecular gas \citep{Lis12,LisPet16} where n(e)/n(H) 
$\approx 2-3 \times 10^{-4}$ as the result of photoionization of carbon and a 
somewhat smaller contribution from cosmic-ray ionization of atomic hydrogen 
\citep{DraBook}.  The electron abundance  declines at higher density as the
H\p\ fraction decreases but n(e) remains high until C\p\ recombines to CO 
\citep{Lis11}.  In this case
\Whcop\ $\propto \int n(H) n(\hcop) dL \propto \int n(e) n(\hcop) dL$.
However, CO is formed from the thermal recombination of \hcop\ with electrons
so \Whcop\ is also proportional to the column-averaged CO formation rate.  
With \WCO\ $\propto$ N(CO), the W$_\hcop$/\WCO\ brightness ratio is equivalent
to taking the ratio of the column-averaged CO formation rate to the resulting 
CO column density.  When this is larger, a higher formation rate is being required 
to produce a given amount of CO, so the destruction rate must be larger.  Viewed 
in this way, the variation in \Whcop\ at \WCO\ = 6-12 K-\kms\ in the leftmost panel 
of Figure 8 suggests a wide range in CO photodestruction rates.

Viewed another way,  
\Whcop/\WCO\ $\propto$ $<$n(H)$>$N(\hcop)/\WCO\ $\propto$ $<$n(H)$>$N(\HH)/N(CO), 
so the \Whcop/\WCO\ brightness ratio also samples the relative CO abundance, testing 
whether the observed CO emission is being produced at reasonable n(H) and N(\HH).
The low-light models that fractionate CO more readily in the right-hand panel 
of Figure 6  will produce a given N(CO) and \WCO\ at smaller 
N(\HH), and so with weaker \hcop\ emission.

These two viewpoints are illustrated with calculations of the \Whcop/\WCO\ brightness
ratio in Figure 10, again showing the data from the left-most panel of
Figure 8, but on logarithmic scales. The models with \G0\  = 1 run through the base 
level of \hcop\ emission with a relatively weak density dependence that confirms 
the consistency of the overall CO formation scheme based on recombination of
the observed relative abundance of \hcop. The models 
reproduce the observed \hcop\ brightness and CO column density at typical 
values of the molecular hydrogen column density. The data toward L121 in the 
field of \zoph\ are reproduced with a stronger radiation field and a higher 
density, along with two of the three datapoints in the blue component of B0528 
and one point in the field around \bll. 

With the weaker illumination needed to enhance \coth\ there is little density 
dependence in the model 
\Whcop/\WCO\ ratios and one curve is plotted for n(H) $= 256 \pccc$ covering 
the density range n(H) = 128 - 512 $\pccc$ at \G0\ =1/2: the reddening values
given along the relevant curve scale inversely with n(H).  As indicated in Figure 
10, and as expected, the low-light models produce a given N(CO) or \WCO\ at much 
smaller N(H) and \EBV, and with \Whcop\ weaker by about a factor two 
compared to the models with fuller illumination.  The low-light conditions that 
reproduce N(CO), \WCO\ and N(\coth) at \WCO\ $\la 5$ K-\kms\ 
account for only about one half of the base level of  \hcop\ 
emission \Whcop\ $\approx 0.02-0.03$ K-\kms\ at \WCO\ $\la 5$ K-\kms\
but with correspondingly smaller total column column density and \EBV. 

The CO hotspots near the mm-wave continuum background sources are explained 
as embedded, localized regions perhaps with higher density, but even more
likely with increased shielding, occupying a small fraction of the 
volume of the overall diffuse molecular medium.  In turn, much or most of
the \hcop\ emission arises in other material that makes a small contribution
to the CO abundance and emission.

\subsection{CO hotspots near mm-wave background sources}

Enhanced and faster formation of \HH\ and CO in localized regions of 
greater density and optical shielding are common features of turbulent
models of the HI$\rightarrow$\HH\ transition in diffuse molecular gas
\citep{shetty11,GodFal+14,ValGod+17,BiaNeu+19}.
Most likely the CO hotspots observed
near the background mm-wave continuum sources and shown in Figure 1
are examples of this phenomenon. The similarity of the sizes (10\arcmin)
and peak brightnesses (5 K-\kms) of the hotspots seen in the vicinity
of the mm-wave background sources in Figure 1 suggest that common physical
phenomena and physical conditions are being sampled.  The extraordinarily 
small \WCO/\Wth\ ratio (3:1) observed toward the West peak around B1954+513 
is in marked contrast to the much larger values, typically 20:1, seen in 
the other directions.  This requires further investigation.

As a specific example, consider that the low-light models with n(H) $= 256 \pcc$ 
shown in Figure 10 are at most about 80\% molecular, ie n(\HH) $\la 100\pcc$, and 
produce 1 K-\kms\ $\le$ \WCO $\le$ 5 K-\kms\ at 0.04 $\la$ \EBV $\la$ 0.07 mag.  
With N(H)/\EBV\ $\approx 8\times 10^{21}\pcc$(mag)$^{-1}$ 
\citep{Lis14yEBV,Lis14xEBV,HenDra17} and [C]/[H] $= 1.6\times 10^{-4}$
\citep{SofLau+04},  one would expect  N(C\p) $= 5.3-9.3\times10^{16}\pcc$.
Because the required CO column densities are $1-5\times10^{15}\pcc$ 
there is ample margin for N(CO) to increase, and CO emission to 
brighten to levels \WCO\ = 5 K-\kms, without upsetting the balance 
of the carbon ionization equilibrium that maintains the preponderance 
of carbon in C\p. At a nominal distance of 250 pc as for the field
around B2251+158 \citep{ZucSpe+18}, a region with \EBV\ = 0.05 mag and
n(H)$= 256~\pcc$ has a size N(H)/n(H) $= 4\times 10^{20}\pcc/256\pccc 
= 0.51$ pc, subtending an angular size 8.0\arcmin. This is a typical
size for the regions of enhanced CO emission in Figure 1.

The required N(H) and \EBV\ scale inversely with the density, so the 
implied angular size would vary approximately as 1/n(H)$^2$.
 The surrounding medium could be an order or magnitude more
expansive with a factor $\ga 3$ density contrast that would suffice to
lower the molecular fraction and substantially weaken the CO emission
\citep{Lis17Frac}. These scalings are
important to understanding the presence of the CO hotspots along
lightly-reddened sightlines, in the absence of evidence for
variations in the reddening comparable to the contrasts in \WCO.
It is not necessary to imagine a contraction of the entire medium into
clumps as small as the CO emission regions shown in Figure 1.


\section{Summary and discussion}

\subsection{\cotw\ and \coth}

Once upon a time we mapped \cotw\ emission around the locations of 
a dozen extragalactic mm-wave 
background continuum sources that had been used to survey for the presence 
of molecular gas by observing the J=1-0 line of \hcop\ in absorption. We 
noted the presence of quite strong peak profile-integrated CO J=1-0 emission 
\WCO\ $\ga$ 5 K-\kms\ in regions of very modest reddening \EBV\ = 0.1 - 0.3 
mag (our CO ``hotspots'') when only much weaker CO emission had been seen a 
few arcminutes away toward the continuum sources themselves. 

In this work we followed up by observing \coth, \coei, 
\hcop\ and CS J=2-1 emission at positions of strong CO emission in five of these 
previously-studied regions and we mapped a more extended portion of the L121 
cloud south of the bright star \zoph\ that has frequently been used to study 
diffuse molecular gas in optical/UV absorption.  We also gathered comparable 
information from the literature to produce a larger sample 
of observations of the carbon monoxide isotopolgues from diffuse 
molecular gas covering a wider range of values of the profile-integrated
\cotw\ emission brightness \WCO.

Observations of the individual regions observed here were  discussed in 
Section 3  and illustrated in Figures 1-5.  The systematics of the carbon 
monoxide isotopologues were discussed in Section 4 and summarized in Figure 6.
As shown in Figure 6 there is a gradual increase in the \Wth/\WCO\
isotopologic integrated brightness ratio from 
${1/40} \le$ \Wth/\WCO\ $\le {1/20}$ at 0.8 K-\kms\ $\la$ \WCO\ $\la 4$ K-\kms\ 
to higher 
values 1:15 - 1:5 with much larger scatter (factor 5) at larger \WCO. 
Carbon monoxide column densities  and line brightnesses are largely 
synonymous, with \WCO\ $\approx$ N(CO)/$(10^{15}\pcc)$ for \WCO\ 
$\la  5$  K-\kms\ and N(CO) $\la 5\times 10^{15}\pcc$ at densities 
n(H) $\ga 100 \pcc$ for all isotopologues. The implication, made clear 
in the right side panel in Figure 6 showing both ratios for both the
column density and intensity ratios,
is that the isotopologic intensity ratios observed at \WCO\ $\la 5$ 
K-\kms\ are very close to the ratio of column densities even when 
comparatively large values 1/40 $\la$ \Wth/\WCO\ $\la$ 1/15 are observed.
Departures from the intrinsic ratio $^{13}C$/$^{12}C$ = 1/62 result
from isotopic C\p\ exchange that dominates selective photodissociation
across the entire observed range in \WCO.

The relationship between CO brightness and column density flattens 
for \WCO\ $\ga$ 5 K-\kms\ where 
N(CO)/\WCO\ $> 10^{15}\pcc$ (K-\kms)$^{-1}$, but substantial
enhancement of \coth\ persists over the entire observed range of \WCO,
and whenever C\p\ is the dominant form of gas-phase carbon (and not
otherwise).  
Ratios N(\coth)/N(\cotw) as small as 1/60 are expected only at very
very small N(CO) where there is no self-shielding, in regions of very
high UV illumination, or when the gas phase carbon has fully recombined 
from  C\p\ to CO. The only sure way to avoid 
factor f $> 1$ enhancement of \coth\ is to be certain that the fraction 
of free gas-phase $^{12}C$ in \cotw\ exceeds 1/f. Of course it is 
pointless to derive N(CO) if the ultimate goal is to determine N(\HH) 
when the relative CO abundance N(CO)/N(\HH) is as small and variable as
it is in diffuse molecular gas.

\subsection{\coei}

\coei\ is observed in UV absorption to be depleted by a factor three below 
the value N(\coei):N(\cotw) = 1:520 corresponding to the intrinsic isotopic 
relative abundance ratio $^{18}$O/$^{12}$C.  We  obtained limits 
\Wei/\Wth\ $< 1/40 - 1/20$ along a handful of sightlines with bright \coth\
lines, vs an intrisic ratio $^{18}$O/$^{13}C = 1/8.4$, see Figure 7.  
 Selective photodissociation of \coei\ is present in our data but
the gas is also enhanced in \coth\ and the emission 
observations did not contrain the depletion of \coei\  
at the levels detected in UV absorption despite the very low brightness
levels to which we chased the \coei\ emission.  Nonetheless high 
isotopologic brightness ratios \Wth/\Wei\ are expected to be a hallmark
of emission from diffuse molecular gas.

\subsection{CS J=2-1 vs. CO and \hcop}

In Figure 8 we showed that CS J=2-1 emission abruptly brightens from
0.03 - 0.04 K-\kms\ to 0.2 K-\kms\ over a narrow 
range of CO brightness \WCO\ = 10-12 K-\kms\ and is bimodal with respect 
to the brightness of \hcop\ with a separation of a factor 4.  As noted
in Section 5.2, The J=2-1 line
of CS and the J=1-0 line of \hcop\ are both in the weak excitation limit 
where  \WCS\ and \Whcop\ are proportional to 
N(CS) and N(\hcop), respectively, so the bimodality of the brightnesses 
represents bimodality of the N(CS)/N(\hcop) ratio with values 5 and 1.25,
as indicated in Figure 9.  Bistable chemical solutions for sulfur-bearing
molecules are now well-studied for dense gas but are not expected in
diffuse molecular gas where the electron fraction is high.

\subsection{\hcop\ and CO}

Comparisons between \WCO\ and \Whcop\ are shown in Figure 8 at left, and,
on a logarithmic scale with model results superimposed, in Figure 10. 
As discussed in Section 5.2, \hcop\ plays a special dual role as progenitor 
of CO via thermal recombination with ambient electrons and as an \HH-tracer 
with fixed relative abundance
N(\hcop)/N(\HH) $= 3\times 10^{-9}$.  Given that the \hcop\ emission line 
brightness \Whcop\ $\propto$ N(\hcop) $\propto$ N(\HH) is mostly due to 
rotational excitation by electrons with which \hcop\ also recombines 
to form CO, \Whcop\ traces N(\HH) while
the \Whcop/\WCO\ brightness ratio traces both the CO photodestruction
rate and the CO abundance relative to \HH.  The chemical model that explains
the presence of CO through \hcop\ recombination also accounts for the baseline
level of \hcop\ emission 0.03 K-\kms\ and log(\Whcop/\WCO) $=-2\pm0.3$  
at modest density n(H) $\approx 100\pccc$.  The  peak \hcop\ brightness 
and the scatter in the \Whcop/\WCO\ brightness ratio both increase considerably 
for \WCO\ $\ga$ 10 K-\kms.  We modeled this as arising from some combination
of stronger illumination (increasing the CO photodissociation rate and the 
required column  densities of \HH\ and \hcop\ required to produce 
a given N(CO) and \WCO), and higher density.

The low-light models that strongly fractionate carbon monoxide and enhance 
\coth\ are very efficient at producing \HH\ and CO, and reproduce the 
observed CO column density and brightness \WCO\ $\la$ 5 K-\kms\ at reddening 
well below even 0.1 magnitudes in Figure 10.
The CO hotspots with \WCO\ = 5 K-\kms\ seen in the vicinity of sightlines
toward extragalactic continuum sources probably represent the sort of localized
regions with higher density and increased optical shielding that are seen in
numerical models of turbulent flow in diffuse molecular gas at the 
HI$\rightarrow$\HH\
transition. In Section 5.3 we gave a numerical example showing that the 
column densities required to produce the observed CO emission imply mildly
sub-pc sized regions subtending angular sizes of about 10\arcmin\ 
($256\pcc$/n(H))$^{-2}$ at a distance of 200 pc, as observed.  
These hotspots are embedded in a more broadly-distributed, 
weakly CO-emitting, less fully-molecular medium having density a few
times lower, which is most easily detected 
in \hcop\ absorption while making a noticeable contribution to the \hcop\ 
emission as discussed in Section 5.2.

\subsection{The bright CO emission contribution of diffuse molecular gas}

Although more attention has recently been paid to conditions under which CO 
is not detectable, CO actually radiates exceptionally brightly on a per-molecule 
basis in subthermally-excited diffuse molecular gas. The CO column density 
producing \WCO\ $\approx 1-10$ K-\kms\ in diffuse, partially molecular gas, 
N(CO) $\approx 10^{15}-10^{16}\pcc$ is far smaller than from the colder, 
denser fully-molecular gas in dark clouds, where C\p\ has fully recombined 
to CO at kinetic temperatures of 10-12 K.  At \AV\ = 5 mag and 
N(CO)/N(\HH) $= 8\times10^{-5}$, 
 the same brightness would be produced with 
N(CO) $\approx 5\times 10^{17}\pcc$ in a fully molecular gas. 

For CO, as with other molecules at the threshold of detectability, 
column density is the main determining factor.
Any density capable of fostering a CO chemistry putting 0.1\% or more
of the free gas phase carbon into CO at 1 magnitude extinction 
will put enough energy into the J=1-0 CO line to render its emission 
detectable at a level \WCO\ $\ga 0.1$ K-\kms.  
Detectability at the 1 K-\kms\ level requires N(CO) $= 10^{15}\pcc$ 
for n(H) $\ga 100 \pccc$, while the expected free gas-phase carbon
column density at 1 magnitude of optical extinction is 
$ 4\times10^{17}\pcc$, implying that only 2.5\% of the free gas-phase
carbon need be in CO. For diffuse molecular gas where N(CO) and \WCO\ are 
proxies for each other around the \WCO\ = 1 K-\kms\ threshold of detectability 
in wide-field surveys,  N(\HH)/\WCO\ $\propto$ N(\HH)/N(CO), and it is 
the CO abundance that determines the CO-\HH\ conversion factor and 
the ability of CO to serve as an \HH\ surrogate. 

\subsection{Dense gas tracers also trace diffuse gas}

 Species like \hcop\ and, especially, HCN, having higher dipole moments 
than CO (2-4 Debye vs 0.11 Debye) and higher ``critical densities'' \citep{Shi15} 
are taken to trace ``dense'' gas, ie molecular gas at higher density than that 
traced by the more easily excited and (independently) more ubiquitous CO.
\Whcop/\WCO\ and \WHCN/\WCO\ ratios 0.01 - 0.02 are observed in the disk of 
the Milky Way \citep{Lis95,HelBli97} and other galaxies 
\citep{BigLer+15,JimBig+19}. CS emission is comparably bright in the Milky Way.

Ratios \Whcop/\WCO\ = 1-2\% are also characteristic of our sightlines 
having weak CO emission \WCO\ =  1-2 K-\kms\ and are explained by models of 
modest density n(\HH) $< 100 \pccc$ (Figure 10).  The 
\Whcop/\WCO\ brightness ratio in diffuse gas is an artifact of the CO 
formation chemistry and \Whcop/\WCO\ is higher for weaker \WCO\ in diffuse 
molecular gas because \hcop\ persists under conditions where CO is poorly 
shielded.   Why simillar ratios should be observed  across a wide range 
of supposedly different environoments in the Milky Way and other galaxies
is an open question.

HCN J=1-0 emission is perhaps the gold standard ``dense gas'' tracer
but observations of HCN in diffuse molecular gas are scarce. Unlike \hcop\
or CS, the HCN brightness is divided among three hyperfine components that
are blended in extragalactic but seen separately in emission that is
resolved spatially, as is the case here.  Toward \bll\ Peak 1
\WHCN/\Whcop\ $ = 1.05\pm0.12$ and \WHCN/\WCO\ $= 0.017\pm0.001$ (Table 2 and 
Appendex B). The main HCN hyperfine component was detected 30\arcmin\ south of 
\zoph\ \citep{Lis97} where \Whcop/\WCO\ = 0.019$\pm0.001$ and \WHCN/\WCO\ 
$\approx 0.011\pm0.002$ if the three hyperfine components of HCN emit in the 
LTE ratio 5:9:1.

Thus, 1-2\% \Whcop/\WCO\ and \WHCN/\WCO\ brightness ratios may arise in gas 
having very modest density, n(\HH) $\ga 100 \pccc$ for the better-sampled
\hcop\ data in Figure 10. By contrast CS J=2-1 emission was detected in our 
data only when \WCO\ $\approx$ 10 K-\kms, which is likely to be denser according
 to Figure 10. 
CS J=2-1 emission is less commonly observed now (eg \cite{BigLer+15,JimBig+19})
because it is not covered by the tunings that simultaneously observe CO and the 
high dipole moment species \cch, HCN and \hcop\ with recent broadband $\lambda$3mm 
IRAM receivers.

\subsection{Separating dense and diffuse gas}

Our observations described the qualities of emission from the diffuse 
molecular gas, not the quantity of this gas in the ISM at large. 
\cite{Rom-DuvHey+16} separated the dense and diffuse gas
contributions observed in \cotw,\coth\ and CS across the Milky Way disk, 
concluding that the diffuse gas fraction increased from 10\% at the 
peak of the molecular ring at R $\approx$ R$_0$/2 to 50\% at R = R$_0$, 
the Solar Circle. A comparable diffuse gas fraction at the Solar Circle, 
40\%, was derived by comparing the \cotw\ emissivity measured in galactic
plane surveys with the brightness of emission from diffuse molecular gas observed 
locally at high galactic latitudes, as done here, extrapolated to a face-on view 
through the disk \cite{LisPet+10}. Some 40\% of the emission viewed from outside 
would arise from diffuse molecular gas well outside regions of higher density or 
extinction. 

The ratio of \cotw\ and \coth\ emissivities \ACO/\Ath\ measured in galactic 
plane surveys
{\footnote \ACO, the quantity measured in galactic plane surveys with units of 
K-\kms\ kpc$^{-1}$ \citep{BurGor78} can be related to the mean matter density:
\ACO\ = d\WCO(R)/dR $\propto <$N(\HH)$>/\lambda  \propto <$n(\HH)$>$ 
if N(\HH) $\propto$ \WCO\ via the CO-\HH\ conversion factor and $\lambda$
is the line of sight geometric mean free path between emitting gas parcels 
\citep{LisGer16}}
changes by a factor two or more across the disk of the Milky Way inside 
the Solar Circle \citep{Rom-DuvHey+16,LisBur+84}, 
from \ACO/\Ath\ $\approx$ 5 to \ACO/\Ath $\ga 10$, even as the \Whcop/\WCO\
brightness ratio does not show a comparable change \citep{Lis95,HelBli97}.
Likewise, the two EMPIRE galaxies having factor $\approx 2$ gradients in
\WCO/\Wth\ across their disks (NGC 4254 and the oddball NGC 628) do not have 
pronounced gradients in \Whcop/\WCO\ \citep{JimBig+19}. This is consistent with 
the inference that \Whcop/\WCO\ $\approx 1-2$\% across environments in galaxy 
disks.

 The  \cotw/\coth\ brightness ratio ranges from 10 to 18 with a mean 
 $<$\cotw/\coth$> \approx 13\pm3$ around R = 5 kpc in the 9 EMPIRE galaxies 
\cite{JimBig+19}.  Such high ratios are not seen in the Milky Way except well 
outside the Solar Circle where the diffuse gas fraction is large and the CO-\HH\ 
conversion factor is expected to be be higher owing to lower metallicity 
\citep{BolWol+13}.  The \cotw/\coth\ 
brightness ratio does not vary much across the disks of 6-7 of the 9 EMPIRE 
galaxies, even as 2 of 3 of those galaxies lacking gradients in \WCO/\Wth\ 
show the \coth/\coei\ brightness ratio increasing outward from 
$\approx 6-8$ to $\ga 20$ \citep{JimMar+17}. There is no comparable information 
on the variation of the \coth/\coei\ brightness ratio across the Milky Way disk.  
The constant Milky Way value 8.3 for the \coth/\coei\ brightness ratio quoted by 
\cite{JimMar+17} was determined \citep{WouHen+08} by observing CO isotopoligic 
emission toward the CO-brightest and densest parts of star-forming regions like W49 
and W3 and was only intended to find the intrinsic isotopic abundance ratio 
[$^{13}$O]/[$^{18}$O], not the brightness ratio in the Milky Way disk ISM.

High \cotw/\coth\ brightness ratios and \coth/\coei\ brightness ratio
gradients suggest a prominent role for diffuse molecular gas in external
galaxy disks and, as noted in Section 6.7, such gas emits with \WHCN/\WCO\ 
and \Whcop/\WCO\ brightness ratios of 1-2\% that are characteristic 
of external galaxy disks.  Dense  gas fractions determined from the 
emission of so-called dense gas tracers should be corrected for this effect.
 
\acknowledgments

I thank the staff of the ARO, Dr. Lucy Ziurys (former ARO Director) and especially 
the telescope operators at the ARO Kitt Peak 12m telescope for their support.  
Comments by the anonymous referee helped to clarify my intentions for this work.

The Kitt Peak 12-Meter Radio Telescope is operated by the Arizona Radio 
Observatory (ARO), which is part of Steward Observatory at the University 
of Arizona.  The State of Arizona provides major financial support for ARO 
operations; aditional funding comes  from the National Science Foundation 
Major Research Instrumentation (MRI) grant to the University 
of Arizona for Development of a State-of-the-Art Multiband Receiver for 
Arizona Radio Observatory's New ALMA Antenna (PI Ziurys AST-1531366).

The National Radio Astronomy Observatory is a facility of the National Science
Foundation operated under cooperative agreement by Associated Universities, Inc.
This manuscript was largely finished while the author was enjoying the hospitality
of the ITU-R and the eponymous Sheraton Hotels in Sharm El-Sheikh and Waikiki.

\facility{ARO Kitt Peak 12m}
\software{DRAWSPEC}






\appendix

\section{Data}

Profile integrals for the new emission data presented here are given in Table A.1

\begin{table*}
\caption{Profile integrals for new data presented here}

{
\begin{tabular}{lcccccccccc}
\hline
Source & RA  & Decl & l & b & Velocity & \cotw\ & \coth & \coei & \hcop\ & CS \\
Source & hh.mmsss  & dd.mmsss & d.dddd & d.dddd & \kms & \Kkms & \Kkms & \Kkms & \Kkms & \Kkms \\
\hline
B0736    & 7.39180 & 1.37046 & 216.9897 & 11.3806 & 4.5..7.5 & 0.78(0.03) & 0.048(0.02)  &  & &  \\ 
B0736Pk & 7.39380 & 1.35246 & 217.0528 & 11.4423 &          & 4.15(0.41) & 0.186(0.017) & && \\ 
B0954 & 9.58472 & 65.33547 & 147.7463 & 43.1316 & 2.5..5.0 & 1.63(0.10) & 0.025(0.010) &  & & \\
B0954WPk &9.59355 & 65.33547 & 145.6831 & 43.2011 & & 6.88(0.22) & 0.900(0.015) & (0.011) & &\\
B1954 &19.55427 & 51.31485 & 85.2984 &11.7569  & 0..2 & 1.59(0.05) & 0.055(0.015) & & & \\
B1954WPk &19.54212 & 51.35485 & 85.2535 & 11.9737 & -1..1 &3.12(0.03) & 1.000(0.011) &(0.008) && \\
B1954EPk &19 .55298 & 51.32085 & 85.28686 & 11.7886 & 0..1.5 & 2.95(0.03) & 0.240 (0.011)& &  (0.011) & (0.011) \\
B2251 &22.53572& 16.08534 & 86.11088 & -38.1838 & -11..-8 & 0.78(0.02) & 0.022(0.007) & & & \\
B2251Pk& 22.53424 & 16.12134 & 86.0856 & -38.1037 & &5.20(0.03) & 0.470(0.015) & (0.010) &(0.0111) & 0.045(0.010) \\
B0528 &5.30564 & 13.31155 & 191.4148 & -11.0534 &0..3&(0.04) &(0.008) & & & \\
B0528A& 5.30537 & 13.28151 & 191.2594 & -11.2337 & &10.05(0.25)& 0.537 (0.009) & &0.272(0.009) & 0.197(0.009)\\
B0528T & 5.31156 & 13.20151 & 191.5765 & -11.0467 && 7.77(0.18) & 0.187(0.022) & &0.177(0.011) & 0.044(0.007)\\
B0528pT &5.30495 & 13.30151 & 191.3768 & -11.0507 & & 0.90(0.21) & $<$0.040 & &0.0564(0.010) & \\
B0528 & 5.30564 & 13.31155 & 191.4148 & -11.0534 &8..12& 2.20(0.04) & 0.107(0.008) & & & \\
B0528A & 5.30537 & 13.28151 & 191.2594 & -11.2337 & &2.22(0.23) & 0.075(0.008) & &0.024(0.008) & (0.008)\\
B0528T & 5.31156 & 13.20151 & 191.5765 & -11.0467 &&6.30(0.16) & 0.808(0.019) & & 0.050(0.008) & (0.008)\\
B0528D& 5.29561 & 13.30151 & 191.5765 & -11.0467& & 10.58(0.20) & 2.104(0.015) & & 0.083(0.011)& 0.033(0.006) \\
B0528N  &5.30591 &13.44351 & 191.1905 & -10.8917 & &8.40(0.31) &1.602(0.101) & & 0.034(0.008)&(0.008) \\
B0528pW &5.30482 & 13.32351 & 191.3402 & -11.0344 & & 6.21(0.20) & 0.804(0.111) & &0.055(0.009) & \\
B0528pT &5.30495 & 13.30151 & 191.3768 & -11.0507 & &5.97(0.19) & 0.920(.120) & &0.074(0.010) & \\
BlLac &22.02433 & 42.16399 & 92.5895 & -10.4411 & -3.5..0.5 & 5.14(0.11) & 0.766(0.022) & (0.010) & &  \\
BlLac Peak1 &22.03010 & 42.2018& 92.6711 & -10.42462 & & 12.38(0.42) & & & 0.208(0.022) & 0.172(0.014) \\
BlLac Peak2 &22.03177 &42.16321 & 92.6740 & -10.5073 & & 9.62(0.44) & & & 0.087(0.018) & 0.089(0.018) \\
BlLac Peak3 &22.03116 & 42.16340 & 92.6591 & -10.4955  & & 10.31(0.42) & & & 0.056(0.013) & 0.107(0.010) \\
BlLac Peak4 & 22.02514 & 42.16019 & 92.6032 & -10.4647 && 11.44(0.35) & & & 0.051(0.010) & 0.058(0.010) \\
L121-Z0& 16.34571 & -12.49496 & 3.9533 & 22.6756& -2.5..-0.7 & 11.42(0.104) & 1.400(0.036) & (0.011) & 0.199(0.024 &0.044(0.014) \\
L121-Z1 &16.34034 & -12.37174 & 3.9917 & 22.9756 && 5.54(0.13) &0.137(0.027) & & 0.159(0.021 & \\
L121-Z2 & 16.35372 & -12.52564 & 4.0144 & 22.5144 & & 9.61(0.15) & 0.822(0.029) &  & 0.157(0.0218) & \\
L121-Z3 &16.34467 & -12.42082 & 4.0366& 22.7866 & &8.97(0.11) & 0.819(0.024) & & 0.099(0.016) & (0.010)\\
L121-Z4 &16.35279 & -12.39184 &4.1866 & 22.6811 & &5.74(0.11) & 0.210(0.020) & & 0.114(0.020) &  \\
L121-Z5 & 16.35172& -12.22472& 4.3972 & 22.8811 & &2.14(0.13) & (0.025) & & 0.0167(0.018) & \\
L121-Z6 & 16.35283 & -12.11216 & 4.5922& 22.9589 & &1.00(0.15) &(0.020) & & 0.037(0.011) & \\
L121-Z7 &16.36532 & -12.21271 & 4.67 00 & 22.5811 & & 11.54(0.16) &0.660(0.020) & & 0.0362(0.012) & \\
L121-Z8 & 16.34506 & -11.43081 & 4.9028& 23.3630 & & 4.48(0.20) & 0.300(0.011) & (0.011)& 0.102(0.017) & \\
L121-Z9 & 16.34491 & -11.35599 & 5.0028 & 23.4389 & &3.67(0.11) & 0.168(0.021) & &0.038(0.014) & \\
L121-Z4  &16.35279 & -12.39184 &4.1866 & 22.6811 & -0.7..1.7 & 9.34(0.11) & 1.161(0.032) &  & 0.089(0.018) & (0.011) \\
L121-Z5  & 16.35172& -12.22472& 4.3972 & 22.8811 & & 5.97(0.13) & 0.312(0.031) & & 0.080(0.018) & \\
L121-Z6 & 16.35283 & -12.11216 & 4.5922& 22.9589& &7.75(0.16) & 0.587(0.022) & & 0.076(0.010) & \\
L121-Z8 & 16.34506 & -11.43081 & 4.9028& 23.3630 & & 12.02(0.14) & 1.381(0.042) & (0.012)& 0.102(0.017) & \\
L121-Z9  & 16.34491 & -11.35599 & 5.0028 & 23.4389 & & 13.91(0.12) & 1.364(0.028) & &0.120(0.015) & \\
\hline
\end{tabular}}
\end{table*}

\section{An extended inventory of emission near \bll}

Figure B.1 shows emission profiles toward Peak 1 near \bll\ including
HCN, HNC and C$_2$H that are not discussed in the main text.
HCN and HNC column densities measured in absorption in diffuse gas are
discussed \cite{LisLuc01} and observations of C$_2$H and oher
small hydrocarbons are discussed in \cite{LucLis00C2H} and 
\cite{LisSon+12}. The profile integrals are 0.218$\pm$0.007 K-\kms\
for HCN (all three components), 0.067$\pm$0.009 K-\kms\ for HNC, and
 0.017$\pm$0.009 K-\kms\ for \cch.

\begin{figure}
\includegraphics[height=8.7cm]{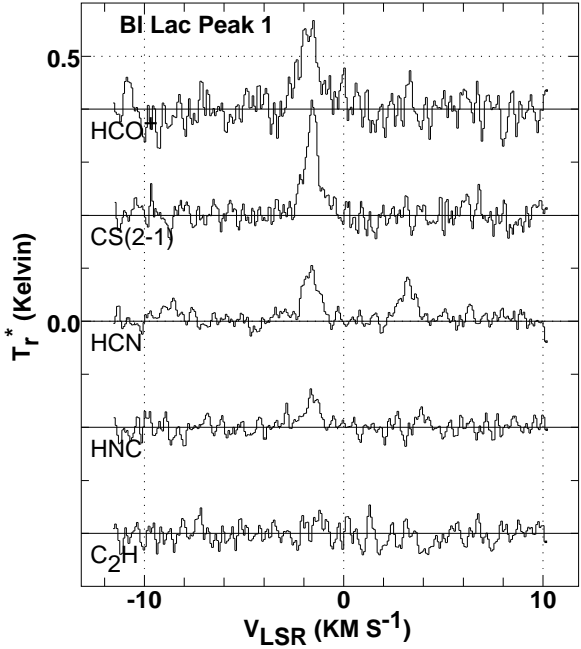}
\caption{A fuller inventory of molecular emission toward Peak 1 near \bll.}
\end{figure}

\section{N(CS) and \hcop\  observed in absorption}

Figure C.1 shows CS and \hcop\ column densities derived in absorption, 
with data as in Figure 4 of \cite{LucLis02} but using an updated permanent 
dipole moment  $\mu_{\rm D} = 3.92$ Debye \citep{MouRed+12} to calculate 
N(\hcop), instead of  4.07 Debye.  The ensemble-averaged
CS/\hcop\ abundance ratio is $<$N(CS)$>$/$<$N(\hcop)$>$ = 1.27 
and the mean CS/\hcop\ ratio is $<$N(CS)/N(\hcop)$>$ = 1.66$\pm$1.32.
Although there is one datapoint with \WCS/\Whcop\ $\approx$ 5
in Figure C.2, the bimodality seen in CS and \hcop\ emission in Figure 
8 is not present in the absorption sample, consistent with the weak 
CO emission that characterizes the sightlines toward the continuum sources.

\begin{figure}
\includegraphics[height=8.7cm]{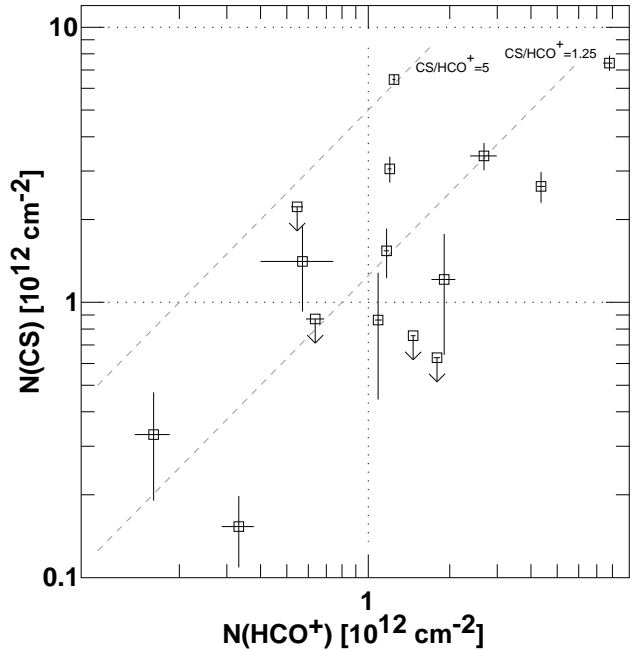}
\caption{CS and \hcop\ column densities measured in absorption, see Appendix C}
\end{figure}

%

\bibliographystyle{aasjournal} 

\end{document}